\documentclass[final]{IEEEtran}
\IEEEoverridecommandlockouts
\usepackage{cite}
\usepackage{amsmath,amssymb,amsfonts}
\usepackage{algorithmic}
\usepackage{graphicx}
\usepackage{textcomp}
\usepackage{xcolor}
\usepackage{bm,soul}
\usepackage{comment}
\usepackage{tikz}
\usepackage{pgfplots}
\usepackage{float}
\usepackage{color}
\usepackage{soul}
\usepackage{mathtools}
\usepackage{setspace}
\usepackage{cite}
\usepackage{scalerel}
\usepackage{comment}
\usepackage{mathtools}
\usepackage{subfigure}
\usepackage[acronym,shortcuts]{glossaries}
\usepackage{todonotes,url}
\usepackage{bbm}
\usepackage{balance}

\def\BibTeX{{\rm B\kern-.05em{\sc i\kern-.025em b}\kern-.08em
    T\kern-.1667em\lower.7ex\hbox{E}\kern-.125emX}}

\newacronym[plural=AoAs,firstplural=angles of arrival (AoAs)]{aoa}{AoA}{angle-of-arrival}
\newacronym[plural=AoDs,firstplural=angles of departure (AoDs)]{aod}{AoD}{angle-of-departure}
\newacronym[plural=CFRs,firstplural= channel frequency responses (CFRs)]{CFR}{CFR}{channel frequency response}
\newacronym[plural=CIRs,firstplural= channel impulse responses (CIRs)]{CIR}{CIR}{channel impulse response}
\newacronym[plural=RNNs,firstplural= recurrent neural networks (RNNs)]{RNN}{RNN}{recurrent neural network}
\newacronym{$k$-NN}{$k$-NN}{k-Nearest Neighbors}
\newacronym{awgn}{AWGN}{additive white Gaussian noise}
\newacronym{BS}{BS}{base station}
\newacronym{CDF}{CDF}{cumulative distribution function}
\newacronym{ce}{CE}{channel estimation}
\newacronym{cearc}{CEARC}{channel estimation with adaptive RIS configuration}
\newacronym{CSI}{CSI}{channel state information}
\newacronym{dft}{DFT}{discrete Fourier transform}
\newacronym{FA}{FA}{false alarm}
\newacronym{lmmse}{LMMSE}{linear minimum mean square error}
\newacronym{LOS}{LOS}{line of sight}
\newacronym{LRT}{LRT}{likelihood ratio test} 
\newacronym{ls}{LS}{least-square}
\newacronym{LSTM}{LSTM}{long short-term memory}
\newacronym{LT}{LT}{likelihood test}
\newacronym{MAP}{MAP}{maximum a posteriori probability}
\newacronym{mbce}{MBCE}{MUSIC-based channel estimation}
\newacronym{mdp}{MDP}{misdetection probability}
\newacronym{MIMO}{MIMO}{multiple-input multiple-output}
\newacronym{miso}{MISO}{multiple-input single-output}
\newacronym{ML}{ML}{machine learning}
\newacronym{mmwave}{mmWave}{millimeter-wave}
\newacronym{mrc}{MRC}{maximum ratio combining}
\newacronym{mse}{MSE}{mean squared error}
\newacronym{music}{MUSIC}{multiple signal classification}
\newacronym{ofdm}{OFDM}{orthogonal frequency division multiplexing}
\newacronym{omp}{OMP}{orthogonal matching pursuit}
\newacronym{parafac}{PARAFAC}{parallel factor}
\newacronym{pdf}{pdf}{probability density function}
\newacronym{PLA}{PLA}{physical layer authentication}
\newacronym{pmf}{PMF}{probability mass function}
\newacronym{RIS}{RIS}{reconfigurable intelligent surface}
\newacronym{simo}{SIMO}{single-input multiple-output}
 \newacronym{snr}{SNR}{signal-to-noise ratio}
\newacronym{ue}{UE}{user equipment}
\newacronym{ULA}{ULA}{uniform linear array}
\newacronym{zf}{ZF}{zero-forcing}

\DeclareMathOperator{\vect}{vec}
\DeclareMathOperator{\diag}{diag}

\newtheorem{remark}{Remark}        
\begin{document}
\sloppy 

\title{Game-Theoretic Analysis of RIS-Assisted Challenge-Response PLA -- \\ Part I: Perfect Channel Knowledge
\thanks{This work was partially supported by European Union (EU) COST Action CA22168—Physical Layer Security for Trustworthy and Resilient 6G Systems (6G-PHYSEC).}
}

\author{Anna V. Guglielmi$^{1}$, Linda Senigagliesi$^{2}$, Marco Baldi$^{3}$, and Stefano Tomasin$^{1}$ \\ \small
\IEEEauthorblockA{$^{1}$University of Padova, Italy,
  $^{2}$ETIS UMR 8051, CYU, ENSEA, CNRS, Cergy, France,\\
 $^{3}$Università Politecnica delle Marche, Ancona, Italy\\
	 email: annavaleria.guglielmi@unipd.it, linda.senigagliesi@ensea.fr,  m.baldi@univpm.it, stefano.tomasin@unipd.it
       }
       }

\maketitle

\begin{abstract}
We consider a physical layer authentication scenario in which a legitimate user (Alice) communicates with an authenticator (Bob) through a reconfigurable intelligent surface (RIS), controlled by Bob. An attacker (Trudy) aims to impersonate Alice by suitably precoding her messages and still transmitting them through the RIS. This two-part study analyzes this attack-defense interaction between Bob and Trudy under perfect (Part I) and partial (Part II) channel knowledge by all the agents. Part I formulates the joint RIS selection and precoder design under perfect channel knowledge as a finite zero-sum game, establishing exact channel subspace conditions for attack indistinguishability alongside equilibrium mixed strategies. Part II extends the defense to incomplete channel information, where each agent observes only its own channel, by formulating a finite-action Bayesian game and characterizing the resulting Bayesian Nash equilibrium. Simulations in geometric and 3GPP TR 38.901 UMi-NLOS channels reveal that authentication performance is governed by available spatial degrees of freedom, channel correlation, and operating signal-to-noise ratio. The proposed equilibrium strategies effectively balance communication rate and security, outperforming conventional benchmark policies.
\end{abstract}


\section{Introduction}

Authentication is the process by which two communicating users can mutually verify each other's identities, or one of them can verify the other's identity.
Authentication is typically performed using asymmetric cryptographic primitives, which follow the computational security paradigm. Although this solution remains the most practical and widely used option to date, the computational security paradigm is affected by advances in cryptanalysis and the evolution of computing platforms. For example, the advent of quantum computers requires replacing current asymmetric cryptography with post-quantum alternatives, whose increased complexity could limit their use in devices with limited resources~\cite{Lopez2025}.

An approach that complements the cryptographic one, which is the focus of this paper and does not follow the computational security paradigm, is that of \ac{PLA}.
\ac{PLA} has been studied in several contexts, including \ac{ofdm} and \ac{MIMO} communications~\cite{Baracca2012, hou2014physical}, underwater acoustic communications~\cite{diamant2019cooperative}, using several techniques for the test, from Neyman-Pearson ~\cite{maurer2000authentication} to machine learning approaches~\cite{Fang2018}, and using various features of received signals, such as \ac{CFR} and \ac{CIR}, to distinguish a legitimate user from a potential attacker \cite{11003929}. Recently, the \ac{aoa} of the signal has been shown to be a robust feature for \ac{PLA} \cite{Pham2023,Srinivasan2024}. In addition, user classification has been done using both classical statistical approaches and modern tools based on machine learning \cite{Senigagliesi2021}.


\Acp{RIS} shape the propagation environment and may enhance \ac{PLA}. Location-optimized \acp{RIS} improve tag-based authentication accuracy in generic settings \cite{Meng2023} and non-line-of-sight vehicular networks \cite{Shawky2023}. Hybrid \acp{RIS} can estimate channels directly to enable \ac{CFR}-based authentication \cite{Selim2023}, whereas cryptographic-assisted approaches incorporate pre-shared keys alongside \acp{RIS} \cite{Zhang2023}. Furthermore, deep learning approaches leverage convolutional neural networks for \ac{CIR}-based \ac{PLA} to mitigate time-varying channel limitations inherent in classical hypothesis testing \cite{Liu2024}.
Besides enhancing conventional \ac{PLA}, \ac{RIS} reconfigurability enables challenge-response \ac{PLA} (CR-PLA) by randomizing the propagation environment to create \ac{CSI}-based challenge-response pairs \cite{Tomasin2022, tomasin2024analysis, icc2, guglielmi2026multi}. In \cite{tomasin2024analysis}, random \ac{RIS} perturbations around communication-optimal settings are evaluated against \ac{MAP}-based attacks, showing that adversaries with sufficient spatial degrees of freedom and perfect channel knowledge can render attacks fully indistinguishable from legitimate transmissions. To refine this approach, CR-PLA schemes optimize the RIS phase perturbation distribution using its first two trigonometric moments, maximizing the achievable communication rate under target \ac{mdp} constraints for single- and multi-antenna systems \cite{icc2, guglielmi2026multi}. Game-theoretic formulations have also been employed to model the interaction between a legitimate verifier and a spoofing attacker as a channel-based authentication game \cite{Xiao2017GT,Zhou2022GT}, although without considering a reconfigurable propagation environment or a challenge-response mechanism.

In our own prior work~\cite{Senigagliesi25}, we studied the geometric conditions under which an impersonation attack becomes indistinguishable from the legitimate transmission at Bob, for a multi-antenna attacker and a single-antenna legitimate user. By deriving the optimal precoding strategy of a fully-informed attacker, we showed that indistinguishability requires matching angles of arrival at the RIS and proportional path gains between the attacker and Alice's channels, highlighting the role of angular alignment and multipath propagation in determining attack feasibility.

In this two-part work, we consider a RIS-assisted CR-PLA system in which the signals transmitted by both the legitimate user Alice and the multi-antenna attacker Trudy reach the verifier Bob through a \ac{RIS} controlled by Bob. Authentication is performed by comparing the estimated cascaded channel with a previously acquired legitimate reference. Bob can randomize the RIS configuration used as the authentication challenge, while Trudy exploits spatial precoding to minimize the mismatch between her forged cascaded channel and Alice's legitimate one. This Part I assumes that all agents know the realization of all the channels. We model Bob-Trudy interaction as a finite two-player zero-sum game, deriving exact channel subspace conditions for perfect attack indistinguishability under full channel knowledge. By characterizing the resulting equilibrium mixed strategies, this work demonstrates that \ac{RIS} randomization yields substantial security gains when the attacker cannot uniformly replicate the channel, while quantifying the inherent communication rate–security trade-off.  Part II \cite{partII} extends the defense to incomplete channel information, where each agent observes only its own channel. We formulate the Bob-Trudy interaction as a finite-action Bayesian game to establish closed-form geometric subspace conditions for arbitrary multi-antenna, multi-path channels when players rely on statistical knowledge of their opponent's channel. Characterizing the Bayesian Nash equilibrium reveals that incomplete channel state information severely limits the attacker's ability to exploit its spatial degrees of freedom, while confirming that the proposed defense consistently outperforms standard deterministic and uniform baseline policies.

Focusing on this Part I, building on our previous work~\cite{Senigagliesi25}, we first extend the analysis of perfect indistinguishability to a general multiple-antenna and multipath setting. We assume that all agents have perfect knowledge of all the channels and investigate two attack models. In the first, Trudy seeks a single configuration-agnostic precoder that reproduces Alice's channel for every RIS configuration. In the second, an omniscient Trudy knows the current RIS configuration and adapts her precoder accordingly. For both cases, we characterize perfect indistinguishability through subspace-inclusion conditions and relate their feasibility to the number of antennas and propagation paths available to Bob and Trudy.

We then move beyond the analysis of an individual attack configuration and formulate the interaction between Bob and Trudy as a finite two-player zero-sum game under perfect channel knowledge. Bob selects a probability distribution over a finite codebook of RIS configurations to minimize the \ac{mdp}, whereas Trudy selects a probability distribution over a finite codebook of precoders to maximize it. The resulting minimax equilibrium provides the optimal randomized authentication policy for Bob and the corresponding worst-case attack strategy for Trudy.

The framework we propose differs from the schemes in~\cite{tomasin2024analysis} and related RIS-assisted CR-PLA works in two main respects. First, rather than evaluating an attacker against a fixed, a priori RIS randomization policy, we jointly determine the attack and defense policies as mutual best responses. Second, we allow Trudy's spatial resources to vary independently of Bob's, including regimes in which her number of antennas or propagation paths is insufficient to reproduce the effective channel observed by Bob. This makes it possible to identify explicitly when RIS randomization provides an authentication advantage and when the system instead collapses to the indistinguishability regime considered in prior work.

\subsection{Contributions of Part I}

The main contributions of this paper can be summarized as follows:
\begin{itemize}

\item We derive exact deterministic conditions under which Trudy can reproduce Alice's cascaded channel under the assumption of perfect channel knowledge by all agents. We consider both a configuration-agnostic precoder, which remains effective for every RIS configuration, and a configuration-dependent precoder employed by an omniscient attacker.

\item We relate these indistinguishability conditions to the spatial dimensions of the Alice--RIS, Trudy--RIS, and RIS--Bob channels. This analysis clarifies the respective roles of the number of antennas and propagation paths and identifies the full-controllability regime in which RIS randomization cannot provide authentication security.

\item We formulate RIS-assisted CR-PLA under perfect channel knowledge as a finite two-player zero-sum game. Its mixed-strategy equilibrium jointly determines Bob's randomized RIS policy and Trudy's randomized precoding policy and yields the worst-case equilibrium \ac{mdp}.

\item We evaluate the proposed framework under both a geometric multipath channel model and the standardized 3GPP TR~38.901 UMi-NLOS model. The results quantify how the equilibrium security depends on the spatial resources of Bob and Trudy, the channel richness, the SNR, and the target false-alarm probability.

\item We compare the proposed minimax strategy with deterministic, uniform, communication-oriented, and state-of-the-art-inspired baselines. We also investigate how the RIS codebook design affects the tradeoff between authentication security and legitimate communication rate.

\end{itemize}

Unlike approaches that directly optimize the RIS perturbation distribution according to a joint rate--security objective, the present work optimizes Bob's policy exclusively with respect to worst-case authentication security. The resulting communication performance is evaluated a posteriori for different fixed RIS codebooks. 

The remainder of the paper is organized as follows. Section~\ref{sec:mod} introduces the system and channel models and states the assumptions on the attacker. Section~\ref{sec:PLA} describes the considered CR-PLA mechanism and its performance metrics. Section~\ref{sec:exact_ind} derives the perfect indistinguishability conditions for configuration-agnostic and configuration-dependent attacks. Section~\ref{sec:probForm} formulates the interaction between Bob and Trudy as a zero-sum game and characterizes its equilibrium strategies. Section~\ref{sec:res} presents the numerical results under the geometric and 3GPP channel models. Finally, Section~\ref{sec:conclusions} concludes the paper.

\section{System Model}
\label{sec:mod}

\begin{figure}[t!]
\centering
\includegraphics[width=90mm]{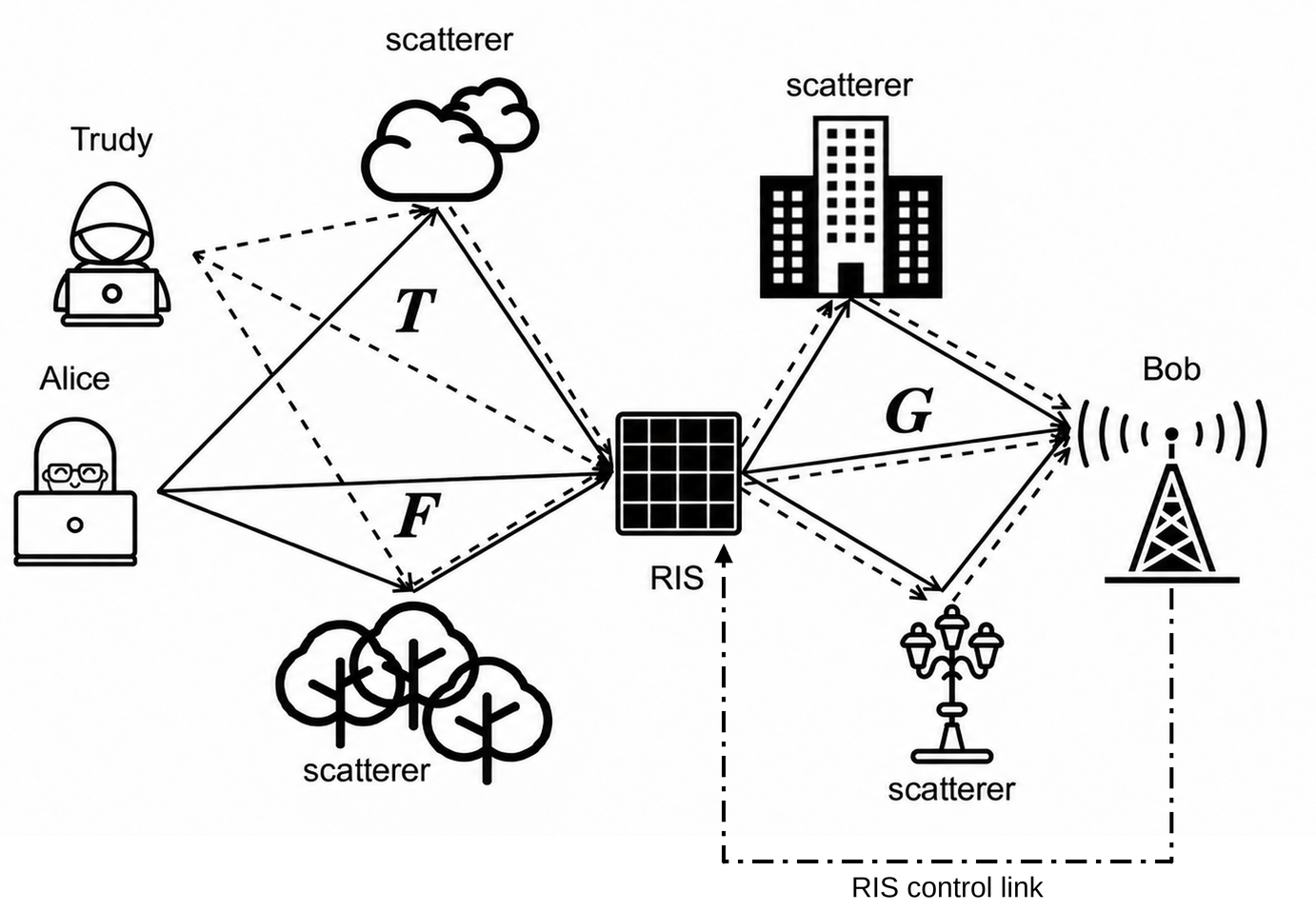}
\caption{System model.}
\label{fig:sys_mod}
\end{figure}

We consider the system model shown in Fig.~\ref{fig:sys_mod}, where the \ac{BS} Bob aims to authenticate a \ac{ue} Alice in a \ac{MIMO} communication scenario. Alice is equipped with a \ac{ULA} of $K$ transmit antennas, while Bob employs a \ac{ULA} of $M$ receive antennas. The signal transmitted by Alice reaches Bob through a \ac{RIS}, while a blockage obstructs the direct Alice--Bob link. An attacker device, Trudy, attempts to impersonate Alice by transmitting signals that Bob might mistake for coming from
Alice. Trudy is equipped with a \ac{ULA} of $N_{\rm T}$ transmit antennas. We also assume that the direct Trudy--Bob link is blocked and that all of Trudy's transmissions reach Bob through the RIS. 
Transmissions occur at \ac{mmwave} frequencies. The antenna elements of the ULAs are uniformly spaced by a distance $d=\lambda_c/2$, where $\lambda_c$ is the carrier wavelength.

The \ac{RIS}, controlled by Bob, comprises $N$ reflecting elements spaced by the same distance $d$. The $n$-th RIS element introduces the phase shift
$\omega_n=e^{j\varphi_n}$, $n=0,1,\ldots,N-1$, on the equivalent baseband signal and is assumed to have a unit-amplitude reflection coefficient. The RIS configuration vector and the corresponding configuration matrix are defined as
\begin{equation}
\bm{\omega}
=
\left[
e^{j\varphi_0},\ldots,e^{j\varphi_{N-1}}
\right]^T,
\qquad
\bm{\Omega}
=
\diag{\bm{\omega}}.
\end{equation}
Let
\begin{equation}
\mathbb T^N
\triangleq
\left\{
\bm\omega\in\mathbb C^N:
|\omega_n|=1,\ n=0,\ldots,N-1
\right\}
\end{equation}
denote the set of feasible RIS configurations. 

We denote the baseband-equivalent Alice--RIS channel matrix by $\bm F\in\mathbb C^{N\times K}$ and the RIS--Bob channel matrix by $\bm G\in\mathbb C^{M\times N}$. Thus, the resulting Alice--RIS--Bob cascaded channel is
\begin{equation}
\label{eq:cascadedCh}
\bm H_{\rm ARB}
=
\bm G\bm\Omega\bm F
\in\mathbb C^{M\times K}.
\end{equation}
Alice transmits suitable pilot symbols to allow Bob to estimate the cascaded channel used for authentication. The pilot signals are assumed to be known to Trudy.

We denote the Trudy--RIS channel matrix by $\bm T\in\mathbb C^{N\times N_{\rm T}}$. Trudy's goal is to impersonate Alice by inducing at Bob a received signal that is as similar as possible to the one generated by Alice. To this end, Trudy applies a precoding matrix $\bm Q\in\mathbb C^{N_{\rm T}\times K}$ to the transmitted signal, including the pilot symbols. The resulting Trudy--RIS--Bob cascaded channel is
\begin{equation}
\label{eq:HTR} 
\bm H_{\rm TRB}
=
\bm G\bm\Omega\bm T\bm Q
\in\mathbb C^{M\times K}.
\end{equation}
We assume that the physical channel matrices $\bm F$, $\bm G$, and $\bm T$ remain constant over the time interval required to perform the considered authentication procedure. The cascaded channels may nevertheless vary as a consequence of the RIS configuration selected by Bob.

Using standard vectorization identities, the legitimate cascaded channel can be written as
\begin{equation}
\label{eq:vecH}
\vect(\bm H_{\rm ARB})
=
\vect(\bm G\bm\Omega\bm F)
=
\left(\bm F^T\diamond\bm G\right)\bm\omega,
\end{equation}
whereas the cascaded channel induced by Trudy is
\begin{equation}
\label{eq:vecT}
\vect(\bm H_{\rm TRB})
=
\vect(\bm G\bm\Omega\bm T\bm Q)
=
\left((\bm T\bm Q)^T\diamond\bm G\right)\bm\omega,
\end{equation}
where $\diamond$ denotes the Khatri--Rao product.

\subsection{Channel Model}

In the presence of objects surrounding the transmitter and receiver, the transmitted signal reaches the receiver through multiple propagation paths. At mmWave frequencies, only a limited number of paths typically carry a significant amount of power; therefore, we employ a geometric channel model.

For a generic ULA with $D$ elements and spatial angle $\theta$, we define the normalized array-response vector as
\begin{equation}
\label{eq:2}
\bm e_D(\theta)
=
\frac{1}{\sqrt D}
\left[
1,
e^{-j\frac{2\pi}{\lambda_c}d\sin\theta},
\ldots,
e^{-j(D-1)\frac{2\pi}{\lambda_c}d\sin\theta}
\right]^T.
\end{equation}
For a generic channel with $L$ paths and angle vector $\bm\theta=[\theta_1,\ldots,\theta_L]^T$, we define the corresponding array-response matrix as
\begin{equation}
\bm E_D(\bm\theta)
=
\left[
\bm e_D(\theta_1),\ldots,\bm e_D(\theta_L)
\right]
\in\mathbb C^{D\times L}.
\end{equation}

Let $L_F$ denote the number of propagation paths between Alice and the RIS. For the $l$-th path, $\phi_{F,l}$, $\theta_{F,l}$, and $\gamma_{F,l}$ denote the \ac{aod} at Alice, the \ac{aoa} at the RIS, and the complex path gain, respectively, for $l=1,\ldots,L_F$. We define $
\bm\phi_F
=
[\phi_{F,1},\ldots,\phi_{F,L_F}]^T,
$, $
\bm\theta_F
=
[\theta_{F,1},\ldots,\theta_{F,L_F}]^T
$,
and
$
\bm\Gamma_F
=
\diag\left(
[\gamma_{F,1},\ldots,\gamma_{F,L_F}]^T
\right)$. Matrices
$\bm E_K(\bm\phi_F)\in\mathbb C^{K\times L_F}$ and
$\bm E_N(\bm\theta_F)\in\mathbb C^{N\times L_F}$
collect the transmit and receive array responses, respectively, while
$\bm\Gamma_F\in\mathbb C^{L_F\times L_F}$ contains the complex path gains. The baseband-equivalent Alice--RIS channel matrix is modeled as~\cite{masood2023inductive}
\begin{align}
\label{eq:3_mod}
\bm F
&=
\sqrt{\frac{KN}{L_F}}
\sum_{l=1}^{L_F}
\gamma_{F,l}
\bm e_N(\theta_{F,l})
\bm e_K^H(\phi_{F,l})
\\
&=
\sqrt{\frac{KN}{L_F}}\,
\bm E_N(\bm\theta_F)
\bm\Gamma_F
\bm E_K^H(\bm\phi_F)
\in\mathbb C^{N\times K}.
\end{align}

Let $L_G$ denote the number of propagation paths between the RIS and Bob. The RIS--Bob channel matrix is modeled as
\begin{equation}
\label{eq:G}
\bm G
=
\sqrt{\frac{MN}{L_G}} \bm E_M(\bm\theta_G)
\bm\Gamma_G
\bm E_N^H(\bm\phi_G)
\in\mathbb C^{M\times N},
\end{equation}
where
$\bm\theta_G\in\mathbb R^{L_G}$ and
$\bm\phi_G\in\mathbb R^{L_G}$
collect the \acp{aoa} at Bob and the \acp{aod} from the RIS, respectively, while
$\bm\Gamma_G\in\mathbb C^{L_G\times L_G}$
is the diagonal matrix containing the corresponding complex path gains.

Similarly, let $L_T$ denote the number of propagation paths between Trudy and the RIS. The Trudy--RIS channel matrix is modeled as
\begin{equation}
\label{eq:6}
\bm T
=
\sqrt{\frac{N_T N}{L_T}} \bm E_N(\bm\theta_T)
\bm\Gamma_T
\bm E_{N_{\rm T}}^H(\bm\phi_T)
\in\mathbb C^{N\times N_{\rm T}},
\end{equation}
where
$\bm\theta_T\in\mathbb R^{L_T}$ and
$\bm\phi_T\in\mathbb R^{L_T}$
collect the \acp{aoa} at the RIS and the \acp{aod} from Trudy, respectively, while
$\bm\Gamma_T\in\mathbb C^{L_T\times L_T}$
is the diagonal matrix containing the corresponding complex path gains.

\subsection{Attacker Model and Assumptions}
\label{Trudy_assumptions}

\paragraph*{Attacker Assumptions} Trudy is an active attacker that attempts to impersonate Alice by transmitting signals that, once received by Bob, resemble the pilot signals transmitted by Alice as closely as possible.
The pilot matrix employed by Alice is assumed to be publicly known and can therefore be reproduced by Trudy. By exploiting her $N_{\rm T}$ transmit antennas, Trudy applies a precoding matrix
$\bm Q\in\mathbb C^{N_{\rm T}\times K}$ to the legitimate pilot matrix, with the objective of making the cascaded channel observed by Bob as close as possible to Alice's legitimate cascaded channel. 
We impose no transmit-power constraint on Trudy's precoder. This assumption allows us to characterize the feasibility of channel emulation without restricting the power required to implement it. The finite-game analysis subsequently restricts
Trudy's choices to the specified codebook $\mathcal C_{\bm Q}$.

\paragraph*{Perfect Channel Knowledge Assumption} We consider a worst-case perfect-channel-knowledge setting. In particular, Trudy is assumed to have instantaneous and error-free knowledge of the channel matrices $\bm F$, $\bm T$, and $\bm G$, as well as of the set of RIS configurations that Bob may employ and of the corresponding randomization policy. Bob does not require knowledge of $\bm T$ to perform the authentication test. However, under the considered perfect-knowledge design, Bob is assumed to know $\bm F$, $\bm T$, and $\bm G$ when optimizing his RIS configuration policy. These assumptions isolate the security gain provided by RIS configuration randomization from channel-estimation uncertainty and provide a conservative assessment of the authentication mechanism.

\paragraph*{Unknown Strategies} Although Trudy knows the set of possible RIS configurations and Bob's randomization policy, she does not observe the current RIS configuration selected for the current authentication attempt before choosing her precoder: Trudy must select her transmission strategy without knowing its current realization. In turn, Alice does not know the current precoder selected by Trudy for transmission.

\section{Physical Layer Authentication Mechanism}
\label{sec:PLA}

We consider a CR-PLA mechanism in which Bob aims at deciding whether a received pilot signal has been transmitted by the legitimate user Alice or by the attacker Trudy. The challenge is represented by the RIS configuration selected by Bob, while the response is the
cascaded channel estimate observed through the Alice--RIS--Bob or Trudy--RIS--Bob link. Accordingly, Bob performs a test on the two hypotheses
\begin{align}
\mathcal H_0 &: \text{the received signal is transmitted by Alice},\\
\mathcal H_1 &: \text{the received signal is transmitted by Trudy}.
\end{align}

The CR-PLA mechanism consists of two phases: an association phase and a verification phase.

\paragraph*{Association Phase} Alice transmits a known pilot matrix $\bm S_0 \in \mathbb{C}^{K\times T_p}$ of a duration of $T_p$ samples, for several RIS configurations. Bob estimates the channel for each configuration and then possibly infers the channel for other RIS configurations \cite{9771077,10625161,piana2026brisc}. At the end of this phase, Bob can compute a reference estimate of the legitimate cascaded Alice--RIS--Bob channel for any configuration $\bm \omega$, denoted $\bar{\bm H}(\bm\omega)$. This phase is assumed to be protected by higher-layer mechanisms, or to be performed in a trusted setup, so that $\bar{\bm H}(\bm\omega)$ can be regarded as a reliable fingerprint of Alice's channel under the considered RIS configuration. The association phase must be repeated whenever the legitimate cascaded channel changes significantly.

\paragraph*{Verification Phase} During the {\em verification phase}, Bob selects a RIS configuration $\bm\omega$ according to his authentication policy and receives a signal that is claimed to originate from Alice. The transmitter (Alice or Trudy) transmits a pilot sequence $\bm S_0$ from which Bob obtains the noisy channel estimate $\hat{\bm H}(\bm \omega)$. When Alice is transmitting, we have 
\begin{equation}
\hat{\bm H}(\bm \omega) = \bm H_{\rm ARB}(\bm\omega)
+
\bm N,
\end{equation}
while when Trudy transmits, we have 
\begin{equation}
\hat{\bm H}(\bm \omega) = \bm H_{\rm TRB}(\bm\omega)
+
\bm N,
\end{equation}
where in both cases $\bm N$ is an \ac{awgn} matrix with complex-Gaussian independent entries with zero-mean with variance $\sigma^2_\mathrm{B}$.
Bob compares the current estimate $\hat{\bm H}$ with the
legitimate reference estimate corresponding to the selected RIS configuration. Since Bob does not rely on instantaneous knowledge of Trudy's channel, the authentication test is based only on the distance from Alice's reference channel. According to the likelihood test (LT), considering the AWGN noise on the channel estimate,  Bob computes
\begin{equation}
\zeta(\bm\omega)
=
\left\|
\hat{\bm H}-\bar{\bm H}(\bm\omega)
\right\|^2,
\label{eq:zeta}
\end{equation}
and applies the test function using threshold $\tau$
\begin{equation}
\begin{cases}
\zeta(\bm\omega) \le \tau,
& \hat{\mathcal H}=\mathcal H_0,\\
\zeta(\bm\omega) > \tau,
& \hat{\mathcal H}=\mathcal H_1.
\end{cases}
\label{eq:thresholdtest}
\end{equation}

\paragraph*{Performance Metrics} Two error events may occur. A \ac{FA} occurs when Bob rejects a legitimate transmission from Alice, whereas a misdetection occurs when Bob accepts a forged transmission from Trudy. The corresponding \ac{FA} probability and \ac{mdp} are
\begin{equation}
P_{\rm FA}
=
\Pr\left[
\zeta(\bm\omega) > \tau
\,\middle|\,
\mathcal H_0
\right],
\quad
P_{\rm MD}
=
\Pr\left[
\zeta(\bm\omega) \le \tau
\,\middle|\,
\mathcal H_1
\right].
\label{probdef}
\end{equation}
We will detail in the following whether, beyond noise, other parameters in \eqref{probdef} should be considered random.

\begin{remark}
\label{rem:pmd_pfa}
The error probabilities in \eqref{probdef} in general depend on the different distributions of $\zeta$ under $\mathcal H_0$ and
$\mathcal H_1$. However, whenever the forged cascaded channel coincides with
the legitimate one for the RIS configuration selected by Bob, i.e.,
$\bm H_{\rm TRB}(\bm\omega,\bm Q)
=
\bm H_{\rm ARB}(\bm\omega)
$, the received signals under the two hypotheses become statistically
equivalent. Hence
$\zeta(\bm\omega)\,|\,\mathcal H_1 \sim \zeta(\bm\omega)\,|\,\mathcal H_0$
and, for any threshold $\tau$, $P_{\rm MD}
=
\Pr\left[\zeta(\bm\omega)\le\tau\,\middle|\,\mathcal H_0\right]
=
1-P_{\rm FA}$.
In this regime, the test \eqref{eq:thresholdtest} carries no discriminating
information, and any operating point lies on the line
$P_{\rm MD}=1-P_{\rm FA}$: decreasing $P_{\rm FA}$ by increasing $\tau$
necessarily increases $P_{\rm MD}$ by the same amount. 
\end{remark}

\paragraph*{Misdetection Probability}  In a CR-PLA system, the RIS configuration acts as a secret or randomized challenge controlled by Bob. Therefore, the \ac{mdp} depends not only on Trudy's precoder $\bm Q$, but also on the RIS configuration selected by Bob and, more generally, on Bob's RIS configuration policy. For fixed channel realizations $\bm F$, $\bm G$, and $\bm T$, a fixed RIS configuration $\bm\omega$, and a fixed attack precoder $\bm Q$, we define the conditional \ac{mdp} (note that $\zeta(\bm \omega)$ depends on $\bm Q$) as
\begin{equation}
P_{\rm MD}(\bm\omega,\bm Q;\bm F, \bm G, \bm T)
=
\Pr\left[
\zeta(\bm\omega) \le \tau
\,\middle|\,
\mathcal H_1,\bm F, \bm G, \bm T,\bm\omega,\bm Q
\right],
\end{equation}
where the probability is taken with respect to the receiver noise and the channel-estimation error. Then, if Bob selects the RIS configuration according to a probability distribution $p(\cdot)$, the attack success probability of a fixed precoder $\bm Q$ is obtained by averaging the conditional \ac{mdp} over Bob's randomized challenge, namely
\begin{equation}
\overline{P}_{\rm MD}(p, \bm Q; \bm F, \bm G, \bm T)
=
\mathbb E_{\bm\omega\sim p(\cdot)}
\left[
P_{\rm MD}(\bm\omega,\bm Q;\bm F, \bm G, \bm T)
\right].
\end{equation}

\section{Indistinguishability Conditions}
\label{sec:mismatch}

The \ac{mdp} defined in Section~\ref{sec:PLA} depends on the distribution of the authentication statistic under an impersonation attack. This distribution is affected by the receiver noise, channel-estimation errors, and the deterministic mismatch between the cascaded channel induced by Alice and the one induced by Trudy. In this section, we isolate the deterministic component of this mismatch and characterize the limiting case in which Alice and Trudy become exactly indistinguishable at Bob.

For a given RIS configuration $\bm\omega$ and a given precoding matrix $\bm Q$, the legitimate and forged cascaded channels are given by \eqref{eq:cascadedCh} and \eqref{eq:HTR}. Using the vectorized expressions in \eqref{eq:vecH} and
\eqref{eq:vecT}, we obtain
\begin{equation}\label{eq:vecForm1}
\operatorname{vec}
\left(
\bm H_{\rm ARB}(\bm\omega)
-
\bm H_{\rm TRB}(\bm\omega,\bm Q)
\right)
=
\bm \Delta(\bm Q)\bm\omega,
\end{equation}
where
\begin{equation}\label{eq:vecForm2}
\bm \Delta(\bm Q)
=
\left(\bm F^T \diamond \bm G\right)
-
\left[(\bm T\bm Q)^T \diamond \bm G\right].
\end{equation}

Under $\mathcal H_1$, the channel estimate can be written as
\begin{equation}
\hat{\bm H}
=
\bm H_{\rm TRB}(\bm\omega,\bm Q)
+
\bm{W},
\end{equation}
where $\bm{W}$ collects receiver noise and channel-estimation errors. Hence, assuming that the reference estimate is $\bar{\bm H}(\bm\omega)\simeq \bm H_{\rm ARB}(\bm\omega)$, the authentication statistic under attack is
\begin{equation}
\zeta(\bm\omega)
=
\left\|
\bm H_{\rm TRB}(\bm\omega,\bm Q)
-
\bm H_{\rm ARB}(\bm\omega)
+
\bm{W}
\right\|^2 .
\label{eq:zeta_attack_mismatch}
\end{equation}
From \eqref{eq:vecForm1} and \eqref{eq:vecForm2}, we can state then that the deterministic mismatch $\|\bm\Delta(\bm Q)\bm\omega\|^2$ controls how close the distribution of the authentication statistic under attack is to its distribution under the legitimate hypothesis.

\subsection{Perfect Indistinguishability}
\label{sec:exact_ind}

Perfect indistinguishability is achieved for a given pair $(\bm\omega,\bm Q)$ when
\begin{equation}
\bm\Delta(\bm Q)\bm\omega=\bm 0,
\label{eq:exact_indistinguishability}
\end{equation}
or, equivalently, $\bm G\bm\Omega\left(\bm F-\bm T\bm Q\right)=\bm 0$. For a fixed precoder $\bm Q$, the set of RIS configurations leading to exact channel indistinguishability is
\begin{equation}
\nu_0(\bm Q)
=
\left\{
\bm\omega\in\mathbb T^N:
\bm \Delta(\bm Q)\bm\omega=\bm 0
\right\}
=
\operatorname{Null}(\bm \Delta(\bm Q))\cap \mathbb T^N .
\label{eq:exact_indistinguishable_set}
\end{equation}

Although $\nu_0(\bm Q)$ is not used as the operational security metric, it provides useful structural insight. If Bob selects a configuration in $\nu_0(\bm Q)$, Alice and Trudy induce the same cascaded channel at Bob. The authentication statistics under $\mathcal H_0$ and $\mathcal H_1$ then have the same distribution and Remark~\ref{rem:pmd_pfa} holds.

To gain further insight into the general framework developed above, we now examine two special cases that admit a closed-form characterization of the indistinguishability conditions. The first case considers Trudy seeking a single precoding matrix $\bm Q$ that achieves perfect indistinguishability for \emph{all} RIS configurations simultaneously, i.e., a configuration-agnostic attack. This corresponds to finding a $\bm Q$ such that $\bm\Delta(\bm Q)=\bm 0$, which eliminates the dependence on $\bm\omega$ entirely and represents the strongest possible attack from Trudy's perspective. The second case assumes that Trudy has full knowledge of the instantaneous RIS configuration and can adapt her precoder accordingly, constituting a worst-case omniscient-attacker benchmark. 

\subsection{Perfect Indistinguishability Regardless of RIS Configuration}
Let us first investigate the case in which Trudy aims to impersonate Alice for any RIS configuration $\bm{\omega}$, i.e.,  from \eqref{eq:vecH} and \eqref{eq:vecT}
\begin{equation}
    ((\bm{TQ})^T \diamond \bm{G}) = ((\bm{F})^T \diamond \bm{G}).
    \label{eq:ind}
\end{equation}
This requires Trudy to find an optimal beamforming matrix $\bm{Q}$. It is important to note that this condition holds regardless of the RIS configuration.

Let $\bm g_n$ denote the $n$-th column of $\bm G$ and let $\bm d_n^T$ denote the $n$-th row of $\bm F-\bm T\bm Q$. From~\eqref{eq:ind} we have $    \bm g_n\bm d_n^T=\bm 0,
    \qquad n=0,\ldots,N-1\,$. Assuming that no RIS element is completely disconnected from Bob, i.e., $\bm g_n\neq\bm 0$ for every $n$, the equality reduces to the matrix equation \begin{equation}
    \bm T\bm Q=\bm F\,.
    \label{eq:TQF}
\end{equation}
Hence, determining the optimal precoding matrix $\bm Q$ boils down to solving a linear system. If $\bm T$ is square and invertible, the solution is simply
\begin{equation}
    \bm Q=\bm T^{-1}\bm F\,.
\end{equation}
More generally, the minimum-norm least-squares solution is
\begin{equation}\label{pseudoinv}
    \bm Q=\bm T^\dagger\bm F,
\end{equation}
where $(\cdot)^\dagger$ denotes the Moore--Penrose pseudoinverse. This solution achieves perfect indistinguishability if and only if $\bm F$ lies in the column space of $\bm T$.

In general, since $\bm{T} \in \mathbb{C}^{N \times N_{\rm T}}$, its column space has dimension at most
\begin{equation}
\operatorname{rank}(\bm{T}) \le \min(N, N_{\rm T}).
\label{eq:Trudy_cond}
\end{equation}
Therefore, a solution of \eqref{eq:TQF} exists if and only if the column space of $\bm{F}$ is contained in that of $\bm{T}$, i.e.,\begin{equation}
    \mathcal R(\bm F)\subseteq\mathcal R(\bm T)\,.
    \label{eq:range_condition_agnostic}
\end{equation}

A sufficient condition to guarantee the existence of a solution for any $\bm{F} \in \mathbb{C}^{N \times K}$ is that $\bm{T}$ has full row rank, i.e., $\operatorname{rank}(\bm{T}) = N$, which requires $N_{\rm T} \geq N$. In other words, Trudy must deploy at least as many antennas as the number of RIS elements. In this case, $\mathcal{R}(\bm{T}) = \mathbb{C}^N$, and perfect impersonation is always feasible regardless of the Alice--RIS channel realization.

However, even if this sufficient condition is not met, exact matching may still be possible for a particular Alice--RIS channel. By exploiting the geometric channel model in \eqref{eq:3_mod}, we observe that the column space of $\bm F$ is contained in the subspace spanned by the RIS array-response vectors associated with the AoAs $\bm\theta_F$, i.e.,
\begin{equation}
    \mathcal R(\bm F)
    \subseteq
    \operatorname{span}\!\left(\bm E_N(\bm\theta_F)\right).
\end{equation}
Similarly, from \eqref{eq:6}, the column space of $\bm T$ is contained in the subspace spanned by the RIS array-response vectors associated with $\bm\theta_T$, i.e., $
    \mathcal R(\bm T)
    \subseteq
    \operatorname{span}\!\left(\bm E_N(\bm\theta_T)\right)
$, 
with $    \operatorname{rank}(\bm T)
    \leq
    \min(N,L_T,N_{\rm T})$.
Therefore, a necessary, but in general not sufficient, condition for the existence of a solution is $
    \operatorname{rank}(\bm T)
    \geq
    \operatorname{rank}(\bm F)$,
which implies
\begin{equation}
\min(N,L_T,N_{\rm T})
    \geq
    \operatorname{rank}(\bm F).
\end{equation}
Indeed, compatible subspace dimensions do not by themselves guarantee the inclusion in \eqref{eq:range_condition_agnostic}: the spatial subspace of the Alice--RIS channel must also be contained in the spatial subspace accessible through the Trudy--RIS channel.

Under nondegenerate path gains and full-rank steering matrices,
$    \operatorname{rank}(\bm F)
    =
    \min(N,L_F,K)
$
and
$    \operatorname{rank}(\bm T)
    =
    \min(N,L_T,N_{\rm T})$.
Therefore, in this generic case, the dimensional requirement becomes
\begin{equation}
    \min(N,L_T,N_{\rm T})
    \geq
    \min(N,L_F,K)\,.
\end{equation}
This condition highlights that successful impersonation requires Trudy to possess sufficient spatial degrees of freedom, either in terms of the number of antennas or the number of scatterers.  Nevertheless, it remains a dimensional requirement: exact configuration-agnostic indistinguishability additionally requires the subspace inclusion $\mathcal R(\bm F)\subseteq\mathcal R(\bm T)$.

\paragraph*{Single-Path RIS-Bob channel}

When the RIS-Bob channel $\bm{G}$ consists of a single propagation path, the general expression in \eqref{eq:G} reduces to
\begin{equation}
    \bm G = \gamma_G \bm{e}_M(\theta_G)\bm{e}^H_N(\phi_G).
    \label{eq:Grank1}
\end{equation}
In this case, $\bm{G}$ has rank one. 
For a fixed RIS configuration $\bm\Omega$, the perfect indistinguishability condition \eqref{eq:exact_indistinguishability}
reduces to
\begin{equation}
    \bm e_N^H(\phi_G)
    \bm\Omega
    \bm T\bm Q
    =
    \bm e_N^H(\phi_G)
    \bm\Omega
    \bm F.
    \label{eq:single_path_condition}
\end{equation}
Thus, for a fixed RIS configuration, it is sufficient for $\bm T\bm Q$ and $\bm F$ to have the same projection onto the effective RIS steering vector $\bm\Omega^H\bm e_N(\phi_G)$. Equation~\eqref{eq:single_path_condition} represents a set of $K$ scalar constraints and reduces to a single scalar constraint when $K=1$. Provided that
\begin{equation}
    \bm e_N^H(\phi_G)\bm\Omega\bm T\neq\bm 0\,,
\end{equation}
it can be satisfied even with $N_{\rm T}=1$ antenna.

Intuitively, when only a single path exists between the RIS and Bob, the received signal lacks spatial richness, and Bob is unable to distinguish two sources that induce the same projection along that path, making impersonation significantly easier for Trudy. Importantly, this simplification applies to a fixed RIS configuration. A single precoder that must guarantee indistinguishability for every RIS configuration remains subject to \eqref{eq:TQF}, provided that every column of $\bm G$ is nonzero.

\subsection{Indistinguishability Condition For Omniscient Trudy} We now consider a stronger attack model, in which Trudy knows the current RIS configuration $\bm{\omega}$ and can adapt her precoding matrix $\bm{Q}$ accordingly. This allows Trudy to optimize her attack for each specific $\bm{\omega}$, at the cost of requiring side information about the RIS. This corresponds to the worst-case scenario, where Trudy is assumed to perfectly know all the channels, including her channel to the RIS, the Alice-\ac{RIS} and \ac{RIS}-Bob channel matrices $\bm{F}$ and $\bm{G}$, and the RIS configuration.

When Trudy is equipped with $N_{\rm T}$ antennas, her optimal precoding matrix $\bm{Q}$ for a given RIS configuration $\bm{\omega}$ is obtained by solving the least-squares problem
\begin{equation}
    \bm{Q}^\star = \arg\min_{\bm{Q}} \|\bm{C}_{A} - \bm{C}'_{T}\bm{Q}\|_F^2 
    = (\bm{C}'_{T})^\dagger \bm{C}_{A},
    \label{eq:opt_Q}
\end{equation}
where $\bm{C}_{A} = \bm{G\Omega}\bm{F} \in \mathbb{C}^{M \times K}$, 
$\bm{C}'_{\rm{T}} = \bm{G\Omega}\bm{T} \in \mathbb{C}^{M \times N_{\rm T}}$, and $\bm{Q} \in \mathbb{C}^{N_{\rm T} \times K}$.  $\| \cdot \|_F^2$ denotes the Frobenius norm, which reduces to the Euclidean norm for $K=1$.
Under nondegenerate path gains and full-rank steering matrices and given \eqref{eq:G}, $\bm{G}$ has rank $\operatorname{rank}(\bm{G}) = \min(M, N, L_G)$ and $\bm{T}$ has rank $\operatorname{rank}(\bm{T}) = \min(N_{\rm T}, N, L_T)$, hence $\bm{C}'_{\rm{T}}$ has rank at most $\min(M, N_{\rm T}, L_G, L_T)$. 

The system $\bm C'_{T}\bm Q=\bm C_{A}$ admits an exact solution if and only if
\begin{equation}
    \mathcal R(\bm C_{A})
    \subseteq
    \mathcal R(\bm C'_{T}).
    \label{eq:omniscient_exact_condition}
\end{equation}
This is the exact deterministic indistinguishability condition for the
considered RIS configuration. Since both matrices contain the common left factor $\bm G\bm\Omega$, a sufficient condition that guarantees
\eqref{eq:omniscient_exact_condition} for any $\bm F$ is $    \operatorname{rank}
    \left(\bm G\bm\Omega\bm T\right)
    =
    \operatorname{rank}(\bm G)$.

For generic channel realizations and nondegenerate path parameters, we have (almost surely, a.s.)
\begin{equation}
    \operatorname{rank}
    \left(\bm G\bm\Omega\bm T\right)
    \stackrel{a.s.}{=}
    \min\left\{
    \operatorname{rank}(\bm G),
    \operatorname{rank}(\bm T)
    \right\}.
\end{equation}
Hence, under this generic-channel assumption, the dimensional condition
$
    \operatorname{rank}(\bm T)
    \geq
    \operatorname{rank}(\bm G)
$ guarantees that Trudy spans the entire channel subspace observable by Bob. Using the geometric channel ranks, this condition becomes
\begin{equation}
    \min(N_{\rm T},N,L_T)
    \geq
    \min(M,N,L_G)\,.
    \label{eq:omniscient_generic_condition}
\end{equation}
Assuming that the RIS has more elements than the considered number of paths,
i.e., $N\geq\max(L_G,L_T)$, \eqref{eq:omniscient_generic_condition} simplifies to
\begin{equation}
    \min(N_{\rm T},L_T)
    \geq
    \min(M,L_G)\,.
    \label{eq:omniscient_generic_simplified}
\end{equation}

In other words, under generic channel conditions, indistinguishability is guaranteed when Trudy has sufficient effective spatial degrees of freedom to span the channel subspace observed by Bob. In the common regime $M\geq L_G$, condition~\eqref{eq:omniscient_generic_simplified} reduces to $N_{\rm T}\geq L_G$ and $L_T\geq L_G$.


This is also consistent with \cite{Senigagliesi25}, where we proved that indistinguishability for a single-antenna attacker is achievable when there is a single path between the RIS and Bob. Conversely, when $\min(N_{\rm T},L_T)<\min(M,L_G)$, Trudy cannot generically span the entire RIS--Bob channel subspace. Nevertheless, exact matching for a particular Alice--RIS channel may still occur if the more general subspace condition \eqref{eq:omniscient_exact_condition} is satisfied.

Table~\ref{tab:ind_conditions} summarizes the perfect indistinguishability conditions derived in this section for the considered attack models.

Table~\ref{tab:ind_conditions} summarizes the perfect indistinguishability conditions derived in this section.

\begin{table*}[t]
\centering
\caption{Perfect Indistinguishability Conditions Under Perfect Channel Knowledge.}
\label{tab:ind_conditions}
\footnotesize
\renewcommand{\arraystretch}{1.25}
\setlength{\tabcolsep}{4pt}
\begin{tabular}{|p{2.9cm}|p{4.2cm}|p{3.2cm}|p{4.7cm}|}
\hline
\textbf{Attack model} & \textbf{Exact condition} & \textbf{Sufficient condition} & \textbf{Generic dimensional condition} \\
\hline
Fixed $(\bm\omega,\bm Q)$
& $\bm G\bm\Omega(\bm F-\bm T\bm Q)=\bm 0$
& --
& -- \\
\hline
Config.-agnostic $\bm Q$ ($\forall\bm\omega$, $\bm g_n\neq\bm 0$)
& $\mathcal R(\bm F)\subseteq\mathcal R(\bm T)$
& $\operatorname{rank}(\bm T)=N$
& $\min(N,L_T,N_{\rm T})\geq\min(N,L_F,K)$ (nec.) \\
\hline
Fixed $\bm\omega$, $L_G=1$
& $\bm e_N^H(\phi_G)\bm\Omega(\bm T\bm Q-\bm F)=\bm 0$
& $\bm e_N^H(\phi_G)\bm\Omega\bm T\neq\bm 0$
& $N_{\rm T}\geq 1$ \\
\hline
Omniscient Trudy
& $\mathcal R(\bm G\bm\Omega\bm F)\subseteq\mathcal R(\bm G\bm\Omega\bm T)$
& $\operatorname{rank}(\bm G\bm\Omega\bm T)=\operatorname{rank}(\bm G)$
& $\min(N_{\rm T},L_T)\geq\min(M,L_G)$ ($N\geq\max(L_G,L_T)$) \\
\hline
\end{tabular}
\end{table*}

\section{Zero-Sum Game Formulation}
\label{sec:probForm}

The CR-PLA mechanism introduced above makes the \ac{mdp} the natural security metric of the system. In this section, we formulate the corresponding robust design problem, characterizing the fundamental attack and defense objectives induced by the authentication statistic.

Throughout this section, both Bob and Trudy are assumed to have perfect instantaneous knowledge of all channel matrices. Hence, there is no private channel information, and the interaction between the two players can be modeled as a complete-information zero-sum game. The randomness of the authentication outcome is nevertheless retained, since the \ac{mdp} accounts for receiver noise and channel-estimation errors.

\subsection{Continuous Strategy Space}

Bob controls the RIS configuration vector $\bm\omega\in\mathbb T^N$, while Trudy controls the precoding matrix $\bm Q\in\mathcal Q$, where $\mathcal Q=\mathbb C^{N_{\rm T}\times K}$ is the continuous admissible precoder space, since no transmit-power constraint is imposed. 
For each pair $(\bm\omega,\bm Q)$ and fixed channel realizations $(\bm F,\bm T,\bm G)$, the conditional \ac{mdp} is
\begin{equation}
\begin{split}
P_{\rm MD}(\bm\omega,\bm Q;\bm F,\bm T,\bm G)
=
\Pr\Big[
&\zeta(\bm\omega)\leq\tau
\,\Big|\,
\mathcal H_1,\bm F,\bm T,\bm G,\\
&\bm\omega,\bm Q
\Big]\,,
\end{split}
\label{eq:pmd_conditional}
\end{equation}
where the probability is taken with respect to the receiver noise and the channel-estimation error.

Therefore, Bob aims at selecting the RIS configuration, or a randomized RIS configuration policy, so as to reduce this probability, whereas Trudy aims at selecting a precoder, or a randomized precoding policy, that increases it. Since all channel matrices are fixed and perfectly known to both players, their dependence is omitted in the following for notational simplicity, and
we write $P_{\rm MD}(\bm\omega,\bm Q)$.

Solving this problem directly in the continuous domain is challenging due to the non-convexity of $\mathbb T^N$, the unit-modulus constraints imposed by the RIS, the continuous admissible precoder space $\mathcal Q$, the threshold-based nature of the authentication test, and the coupling between the optimization variables.

\subsection{Finite-Codebook Approximation}

To obtain a tractable formulation, the continuous action spaces of both players are approximated through finite codebooks. In particular, Bob selects the RIS configuration from
\begin{equation}
\mathcal C_\Omega
=
\left\{
\bm\omega_1,\ldots,\bm\omega_{N_\Omega}
\right\}
\subseteq
\mathbb T^N\,,
\end{equation}
whereas Trudy selects the precoding matrix from
\begin{equation}
\mathcal C_Q
=
\left\{
\bm Q_1,\ldots,\bm Q_{N_Q}
\right\}
\subseteq
\mathcal Q\,.
\end{equation}
This yields a finite matrix game that preserves the essential strategic structure of the continuous problem. The construction of $\mathcal C_\Omega$ and $\mathcal C_Q$ is detailed in
Section~\ref{sec:res}.

We frame the resulting optimization problem as a game involving two players: Bob, the defender, who controls the RIS configuration $\bm\omega$, and Trudy, the attacker, who selects a precoding matrix $\bm Q$ to forge the legitimate cascaded channel.

\paragraph*{Actions}
Bob selects a RIS configuration $\bm\omega_i\in\mathcal C_\Omega$, while Trudy selects a precoding matrix $\bm Q_j\in\mathcal C_Q$. The two players select their actions simultaneously.
Therefore, Trudy knows Bob's RIS codebook and randomization policy, but she does not observe the RIS configuration selected for the current authentication attempt before choosing her precoder.

\paragraph*{Payoff Function}
For a given RIS configuration $\bm\omega_i$ and precoder $\bm Q_j$, we define Trudy's payoff as the conditional \ac{mdp}, i.e.,
\begin{equation}
\label{eq:payoff}
u(\bm\omega_i,\bm Q_j)
=
P_{\rm MD}(\bm\omega_i,\bm Q_j)\,.
\end{equation}
This payoff represents the probability that Bob accepts Trudy's forged transmission as legitimate. Trudy aims at maximizing $u$, whereas Bob aims at minimizing the same quantity. Equivalently, Bob's payoff can be defined as $-u$, making the interaction a zero-sum game.

Let
\begin{equation}
\Delta_{N_\Omega}
=
\left\{
\bm o\in\mathbb R_+^{N_\Omega}:
\bm 1^T\bm o=1
\right\}
\end{equation}
and
\begin{equation}
\Delta_{N_Q}
=
\left\{
\bm a\in\mathbb R_+^{N_Q}:
\bm 1^T\bm a=1
\right\}
\end{equation}
denote the probability simplexes over Bob's and Trudy's action sets, respectively. Bob's mixed strategy is denoted by $\bm o=[o_1,\ldots,o_{N_\Omega}]^T\in\Delta_{N_\Omega}$, while Trudy's mixed strategy is denoted by $\bm a=[a_1,\ldots,a_{N_Q}]^T\in\Delta_{N_Q}$.

Let $\bm P\in[0,1]^{N_\Omega\times N_Q}$ be Trudy's payoff matrix, whose entry $p_{ij}$ is the conditional \ac{mdp} obtained when Bob employs $\bm\omega_i$ and Trudy employs $\bm Q_j$, i.e.,
\begin{equation}
p_{ij}
=
P_{\rm MD}(\bm\omega_i,\bm Q_j)\,.
\label{eq:payoff_matrix_entry}
\end{equation}
When Bob and Trudy use the mixed strategies $\bm o$ and $\bm a$, respectively, Trudy's expected payoff is the average \ac{mdp}
\begin{equation}
U(\bm o,\bm a)
=
\bm o^T\bm P\bm a\,.
\label{eq:expected_payoff}
\end{equation}

All equilibrium and worst-case performance claims below refer to the considered finite codebooks. We include $\bm Q_{\rm agn}$ in $\mathcal{C}_Q$ ensuring that the configuration-agnostic least-squares attack is available, but does not establish optimality over the continuous precoder space $\mathcal Q$.

\subsection{Mixed-Strategy Equilibrium}

Since the action sets $\mathcal C_\Omega$ and $\mathcal C_Q$ are finite, and both players are allowed to use mixed strategies, von Neumann's minimax theorem~\cite{Neumann1928} guarantees the existence of a saddle-point equilibrium.

Bob's equilibrium strategy satisfies
\begin{equation}
\label{optZS_bob}
\bm o^\star
\in
\arg\min_{\bm o\in\Delta_{N_\Omega}}
\max_{\bm a\in\Delta_{N_Q}}
U(\bm o,\bm a),
\end{equation}
whereas Trudy's equilibrium strategy satisfies
\begin{equation}
\label{optZS_trudy}
\bm a^\star
\in
\arg\max_{\bm a\in\Delta_{N_Q}}
\min_{\bm o\in\Delta_{N_\Omega}}
U(\bm o,\bm a).
\end{equation}
The minimax theorem ensures that
\begin{equation}
\begin{split}
V_{\rm MD}
&=
\min_{\bm o\in\Delta_{N_\Omega}}
\max_{\bm a\in\Delta_{N_Q}}
U(\bm o,\bm a)
\\
&=
\max_{\bm a\in\Delta_{N_Q}}
\min_{\bm o\in\Delta_{N_\Omega}}
U(\bm o,\bm a)
\\
&=
U(\bm o^\star,\bm a^\star),
\end{split}
\label{optZS}
\end{equation}
where $V_{\rm MD}$ is the unique value of the game.

The resulting strategy profile constitutes a Nash equilibrium. Given Trudy's equilibrium strategy $\bm a^\star$, Bob cannot reduce the expected \ac{mdp} by unilaterally deviating from $\bm o^\star$. Similarly, given $\bm o^\star$, Trudy cannot increase the expected \ac{mdp} by unilaterally deviating from $\bm a^\star$. Equivalently, the saddle-point inequalities
\begin{equation}
U(\bm o^\star,\bm a)
\leq
U(\bm o^\star,\bm a^\star)
\leq
U(\bm o,\bm a^\star)
\end{equation}
hold for every
$\bm o\in\Delta_{N_\Omega}$ and
$\bm a\in\Delta_{N_Q}$.

From a security perspective, $V_{\mathrm{MD}}$ represents
the minimum worst-case MDP that Bob can guarantee when his RIS configurations and Trudy's precoders are restricted to $\mathcal C_\Omega$ and $\mathcal C_Q$,  respectively.

\subsection{Linear-Programming Solution}

Problem~\eqref{optZS} can be solved by exploiting its matrix-game structure. For a fixed strategy $\bm o$ of Bob, Trudy's best response is attained by at least one pure precoder $\bm Q_j$. Therefore,
\begin{equation}
\max_{\bm a\in\Delta_{N_Q}}
\bm o^T\bm P\bm a
=
\max_{j=1,\ldots,N_Q}
\sum_{i=1}^{N_\Omega}
o_i p_{ij}\,.
\end{equation}
Hence, Bob must choose $\bm o$ so as to minimize the largest payoff that Trudy can obtain with any pure attack action. Introducing the auxiliary variable $v$, Bob's problem becomes
\begin{align}
\label{optZS1}
\min_{\bm o,v}\quad
&v
\\
\text{\rm s.t.}\quad
&
\sum_{i=1}^{N_\Omega}
o_i p_{ij}
\leq v,
\qquad
j=1,\ldots,N_Q,
\nonumber\\
&
\sum_{i=1}^{N_\Omega}o_i=1,
\qquad
o_i\geq0,
\qquad
i=1,\ldots,N_\Omega.
\nonumber
\end{align}

Similarly, for a fixed strategy $\bm a$ of Trudy, Bob's best response is attained by at least one pure RIS configuration $\bm\omega_i$. Therefore,
\begin{equation}
\min_{\bm o\in\Delta_{N_\Omega}}
\bm o^T\bm P\bm a
=
\min_{i=1,\ldots,N_\Omega}
\sum_{j=1}^{N_Q}
a_jp_{ij}.
\end{equation}
Thus, Trudy chooses $\bm a$ so as to maximize the smallest payoff that she can guarantee against any pure RIS configuration. Introducing the auxiliary variable $z$, Trudy's problem becomes
\begin{align}
\label{optZS2}
\max_{\bm a,z}\quad
&z
\\
\text{\rm s.t.}\quad
&
\sum_{j=1}^{N_Q}
a_jp_{ij}
\geq z,
\qquad
i=1,\ldots,N_\Omega,
\nonumber\\
&
\sum_{j=1}^{N_Q}a_j=1,
\qquad
a_j\geq0,
\qquad
j=1,\ldots,N_Q.
\nonumber
\end{align}

Problems~\eqref{optZS1} and~\eqref{optZS2} are primal--dual formulations of the same zero-sum matrix game. Strong duality, in accordance with von Neumann's minimax theorem, ensures that the optimal value $v^\star$ found through Bob's problem is equal to the optimal value $z^\star$ found through Trudy's problem, i.e.,
\begin{equation}
v^\star=z^\star=V_{\rm MD}.
\end{equation}
Thus, $V_{\rm MD}$ is the equilibrium \ac{mdp} under perfect channel knowledge and represents the fundamental worst-case vulnerability of the system over the considered finite action codebooks.

\section{Numerical Results}
\label{sec:res}

In this section, we assess the performance of the proposed game-theoretic CR-PLA under the perfect channel knowledge scenario. We first describe the finite action codebooks used by Bob and Trudy and then evaluate the resulting \ac{mdp} $V_{\mathrm{MD}}$ under different system and channel configurations.

Unless otherwise stated, for the geometric channel model, Alice is equipped with a single antenna, i.e., $K=1$, and the RIS comprises $N=32$ reflecting elements with half-wavelength spacing. The Alice--RIS channel contains $L_F=3$ propagation paths. The numbers of antennas at Bob and Trudy are selected from $M\in \{4,8,12,16,20,24,32\}$ and
$N_T\in\{4,8,12,16,20,24,28,32\}$, respectively. The numbers of paths in the Trudy--RIS and RIS--Bob channels are selected from $L_T\in\{1,3,5,7,10\}$ and $L_G\in\{1,3,7,10\}$, respectively. The specific values of $M$, $N_T$, $L_T$, $L_G$, SNR, and $P_{\mathrm{FA}}$ employed in each experiment are reported in the corresponding figure caption. Each reported point is obtained by solving the game for independent channel realizations and subsequently averaging the resulting game values $V_{\mathrm{MD}}$.

Note that we set $\sigma_{\rm B}^{2}=1/\mathrm{SNR}$, with $\mathrm{SNR}$ in linear scale. Under the Gaussian estimation-error model and an exact reference, the normalized test statistic follows a central chi-squared distribution under $\mathcal H_0$ and a noncentral chi-squared distribution under $\mathcal H_1$, both with $2MK$ degrees of freedom~\cite{tomasin2024analysis, guglielmi2026multi}. These distributions are used to set the threshold for the target $P_{\rm FA}$ and evaluate the payoff entries analytically.  
The proposed mixed-strategy equilibrium is compared against the following five benchmarks:
\paragraph*{Uniform Bob + BR Trudy}: Bob randomizes uniformly over the RIS codebook $\mathcal{C}_\Omega$, while Trudy plays her best response over $\mathcal{C}_Q$.

\paragraph*{Pure Bob + BR Trudy}: Bob deterministically employs his best pure RIS configuration, while Trudy plays her best response over $\mathcal{C}_Q$.

\paragraph*{Uniform Bob + Uniform Trudy}: both players randomize uniformly over their respective codebooks, with no best-response computation on either side.

\paragraph*{Comm-optimal Bob + BR Trudy}~\cite{Senigagliesi25}: Bob sets the communication-optimal RIS configuration $\bar{\bm{\Omega}}$, while Trudy uses the corresponding least-squares best-response precoder. Such a precoder achieves perfect impersonation only when the legitimate effective channel belongs to the subspace controllable by Trudy.

\paragraph*{Uniform Bob + Agnostic Trudy}~\cite{tomasin2024analysis}:  Bob randomizes uniformly over its codebook, while Trudy employs the configuration-agnostic least-squares precoder $\bm Q_{\mathrm{agn}}=\bm T^\dagger\bm F$, computed once and reused against every RIS configuration. This choice isolates the effect of the configuration-agnostic attack from that of the RIS codebook design \footnote{Note that this choice does not reproduce the complete scheme of~\cite{tomasin2024analysis}, where Bob generates continuously perturbed RIS configurations around the communication-optimal profile, but rather evaluates its perfect-knowledge attack principle within the discrete RIS action space considered here.} When $\bm T$ has full row rank, $\bm T\bm Q_{\mathrm{agn}}=\bm F$, and the attack recovers the indistinguishability result of~\cite{tomasin2024analysis}. Section~\ref{sec:3gpp_perfect} identifies the regime in which this condition is met.

\subsection{Codebook Design}
\label{subsec:codebDesignOmega}

To provide a structured set of distinct RIS challenges, the codebook $\mathcal{C}_{\Omega}$ contains $N_\Omega=N$ RIS configurations, whose configuration vectors are the columns of an $N\times N$ DFT matrix after removing its normalization factor, i.e., $[\bm\omega_k]_n
e^{-j2\pi nk/N}$, $n, k\in{0,\ldots,N-1}$. 



Since Trudy knows the instantaneous realizations of channels $\bm F$, $\bm T$, and $\bm G$, as well as Bob's codebook $\mathcal C_\Omega$, she constructs her codebook by associating one least-squares precoder with each RIS configuration. Specifically, for the $i$-th RIS configuration $\bm\Omega_i$, she computes
\begin{equation}
\bm Q_i
\left(\bm G\bm\Omega_i\bm T\right)^\dagger
\bm G\bm\Omega_i\bm F,
\qquad
i=1,\ldots,N_\Omega\,.
\label{eq:trudy_codebook_num}
\end{equation}
We additionally include the configuration-agnostic precoder
$\bm{Q}_{\rm agn}=\bm{T}^\dagger \bm{F}$, yielding the attacker codebook $\mathcal C_Q=\{\bm{Q}_1,\ldots,\bm{Q}_{N_\Omega},\bm{Q}_{\rm agn}\}$. Thus, $N_Q=N_\Omega+1$. Unlike the configuration-specific candidates, $\bm{Q}_{\rm agn}$ minimizes the pre-RIS mismatch $\|\bm{F}-\bm{TQ}\|_{\rm F}^2$ independently of the selected RIS configuration. 
Each $\bm Q_i$ minimizes the residual channel mismatch for the corresponding RIS configuration $\bm\Omega_i$, but it is not generally optimal when Bob selects another configuration.
 These precoders are used without power normalization.
When the matrices involved in the pseudoinverse are ill-conditioned, channel emulation may require large transmit power. Consequently, the perfect-indistinguishability conditions established here characterize feasibility without a power constraint; they do not guarantee feasibility under a prescribed transmit-power budget.

Section~\ref{subsec:geoMod} also evaluates an alternative codebook construction in which the RIS configurations are obtained by perturbing the communication-optimal configuration $\bar{\bm\Omega}$. This allows a direct comparison between the DFT-based design and a state-of-the-art-inspired randomization strategy in terms of the resulting rate--security tradeoff.

\subsection{Performance on Geometric Channel}
\label{subsec:geoMod}

We first consider the perfect channel knowledge scenario under the geometric multipath channel model introduced in Section~\ref{sec:mod}. Here, perfect channel knowledge refers exclusively to the instantaneous CSI available to the players. The Alice--RIS and Trudy--RIS channels are generated as distinct channel realizations and are not assumed to coincide. Path gains are independently drawn from $\mathcal{CN}(0,1)$. Arrival and departure angles are independently uniform over $[-\pi/2,\pi/2]$, except for the arrival angles at Bob, which are uniform over $[-\pi/6,\pi/6]$. Unless otherwise specified, the parameters are selected from the sets summarized at the beginning of this section.

\begin{figure}[]
\centering
\includegraphics[width=\columnwidth]{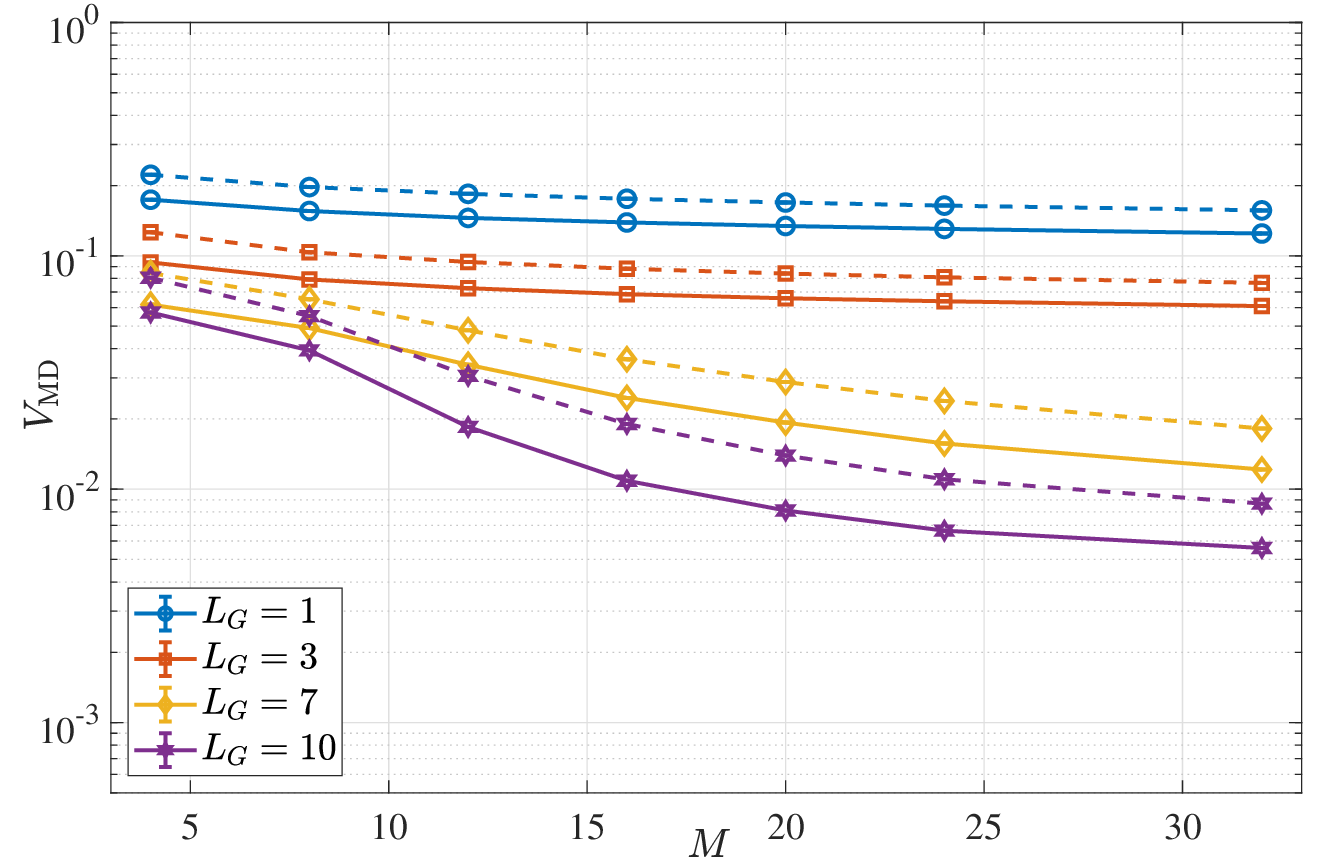}
\caption{
$V_{\mathrm{MD}}$ as a function of the number of antennas at Bob, $M$, for different values of $L_G$, with $L_T=5$, $N_T=16$, and $\mathrm{SNR}=-5$~dB. Solid and dashed lines refer to $P_{\mathrm{FA}}=10^{-1}$ and $P_{\mathrm{FA}}=10^{-2}$, respectively.
}
\label{fig:pmd_m_pfa1}
\end{figure}

\begin{figure}[]
\centering
\includegraphics[width=\columnwidth]{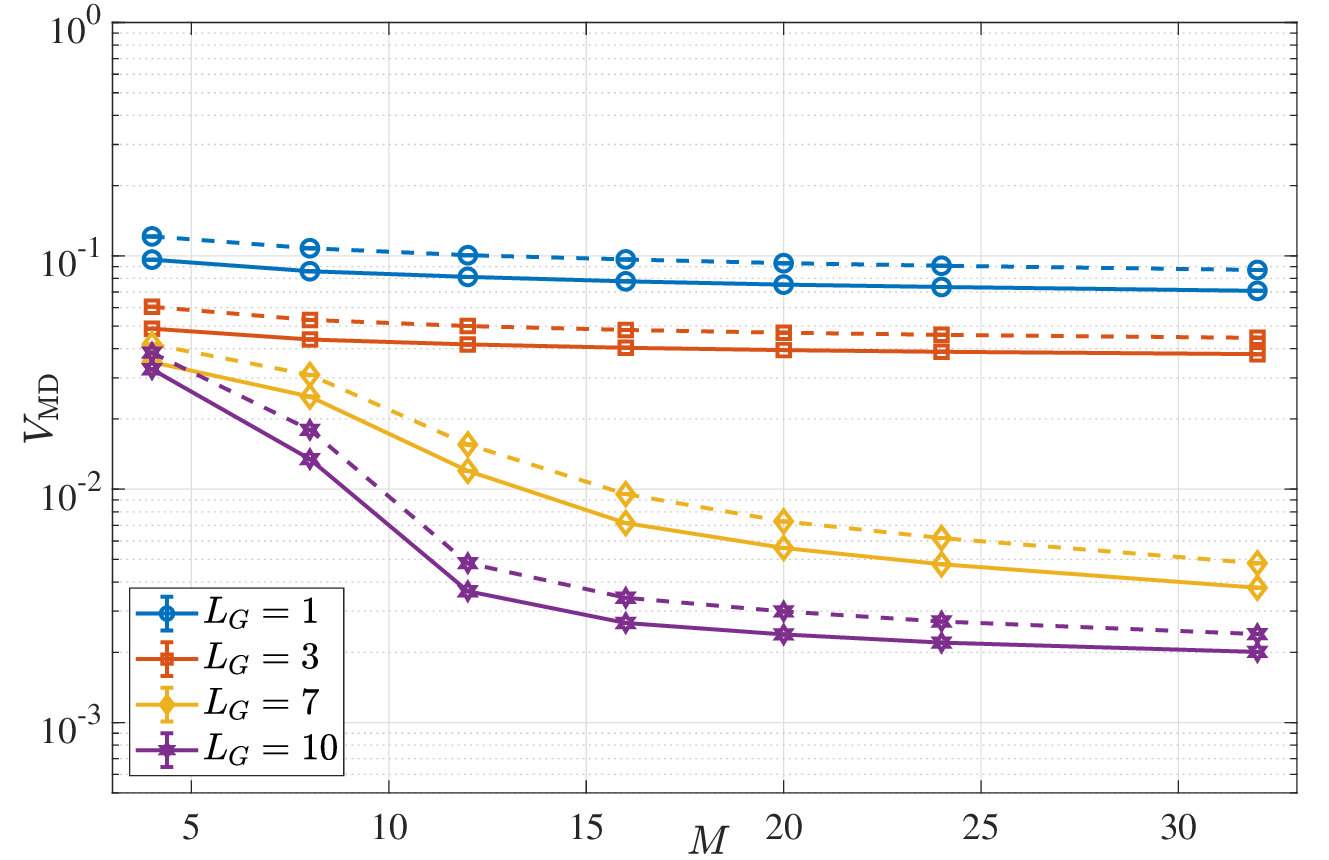}
\caption{
$V_{\mathrm{MD}}$ as a function of the number of antennas at Bob, $M$, for different values of $L_G$, with $L_T=5$, $N_T=16$, and $\mathrm{SNR}=5$~dB. Solid and dashed lines refer to $P_{\mathrm{FA}}=10^{-1}$ and $P_{\mathrm{FA}}=10^{-2}$, respectively.
}
\label{fig:pmd_m_pfa2}
\end{figure}

Figs.~\ref{fig:pmd_m_pfa1} and~\ref{fig:pmd_m_pfa2} show $V_{\mathrm{MD}}$ as a function of $M$, for $N_T=16$ and $L_T=5$, and two values of SNR, namely $-5$ and $5$~dB. The curves correspond to different values of $L_G$, while solid and dashed lines refer to the target false-alarm probabilities $P_{\mathrm{FA}}=10^{-1}$ and $P_{\mathrm{FA}}=10^{-2}$, respectively. The number of paths in the RIS--Bob link has a major impact on $V_{\mathrm{MD}}$. Small values of $L_G$ yield the largest \ac{mdp}, whereas increasing $L_G$ improves Bob's ability to discriminate between Alice and Trudy. Indeed, a richer RIS--Bob channel exposes the legitimate and forged signals over a higher-dimensional effective subspace, which is more difficult for Trudy to reproduce when the rank of her channel is limited by $L_T$ and $N_T$.

The physical layer authentication performance depends heavily on the spatial richness $L_G$ of the RIS--Bob channel relative to Bob's antenna array size $M$. Increasing $M$ reduces the miss detection probability $V_{\mathrm{MD}}$ only when $L_G$ is large, as a small $L_G$ forms the primary performance bottleneck. Detection is enhanced whenever $\min\{M,L_G\}>\min\{N_T,L_T\}$, exceeding Trudy's available spatial dimension.

While a configuration-specific precoder $\bm Q$ achieves the indistinguishability limit $V_{\rm MD}= 1-P_{\mathrm{FA}}$ for a given RIS setup $\bm\Omega$, Bob's randomized policy degrades its attack effectiveness across other configurations. For generic geometric channels, guaranteeing full controllability requires $N_T \geq N$ and $L_T \geq N$, allowing Trudy to obtain the precoder from \eqref{pseudoinv}  and recover the perfect through-RIS attack in~\cite{tomasin2024analysis} as a limiting case.

Finally, higher SNR systematically lowers $V_{\mathrm{MD}}$ because channel-estimation noise is less effective at masking residual channel mismatches. Ultimately, increasing $M$ improves authentication only when the RIS--Bob channel provides sufficient multipath diversity.

\paragraph*{Effect of varying $N_T$ and $L_T$}

\begin{figure}[]
\centering
\includegraphics[width=\columnwidth]{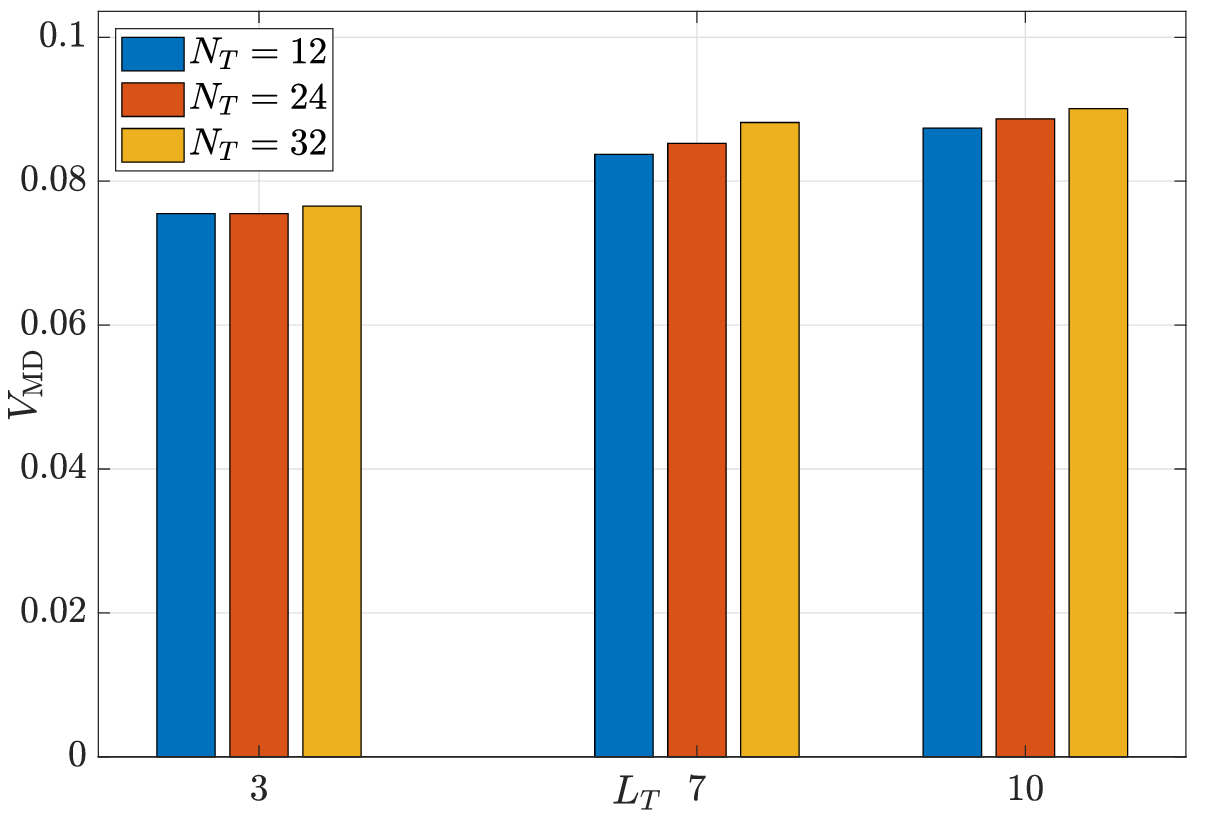}
\caption{
$V_{\mathrm{MD}}$ as a function of $L_T\in{3,7,10}$ and $N_T\in{12,24,32}$, for $M=20$, $L_G=3$, $\mathrm{SNR}=-5$~dB, and $P_{\mathrm{FA}}=10^{-2}$.
}
\label{fig:pmd_m_lg_fixed}
\end{figure}

Fig.~\ref{fig:pmd_m_lg_fixed} shows $V_{\mathrm{MD}}$ for $L_T\in\{3,7,10\}$ and $N_T\in\{12,24,32\}$, with $M=20$, $L_G=3$, $\mathrm{SNR}=-5$~dB, and $P_{\mathrm{FA}}=10^{-2}$. For each fixed value of $L_T$, increasing $N_T$ produces only limited variations in $V_{\mathrm{MD}}$. This behavior is consistent with \eqref{eq:Trudy_cond}.
Since all the considered values of $N_T$ are at least as large as $L_T$, Trudy's effective spatial degrees of freedom are mainly limited by the number of propagation paths rather than by the number of transmit antennas.

Conversely, increasing $L_T$ yields a more evident increase in $V_{\mathrm{MD}}$. A richer Trudy--RIS multipath channel not only enlarges the subspace controllable by Trudy but may also make a precoder designed for one RIS configuration more effective against other configurations. The residual mismatches are consequently reduced, increasing the equilibrium attack success probability.

Thus, once $N_T\geq L_T$, adding further antennas at Trudy provides only marginal benefits, whereas increasing the channel richness may substantially improve the attack. This confirms that the relevant resource is Trudy's effective spatial rank, rather than her number of antennas alone.

\paragraph*{Baselines Comparison}

\begin{figure}[]
\centering
\includegraphics[width=\columnwidth]{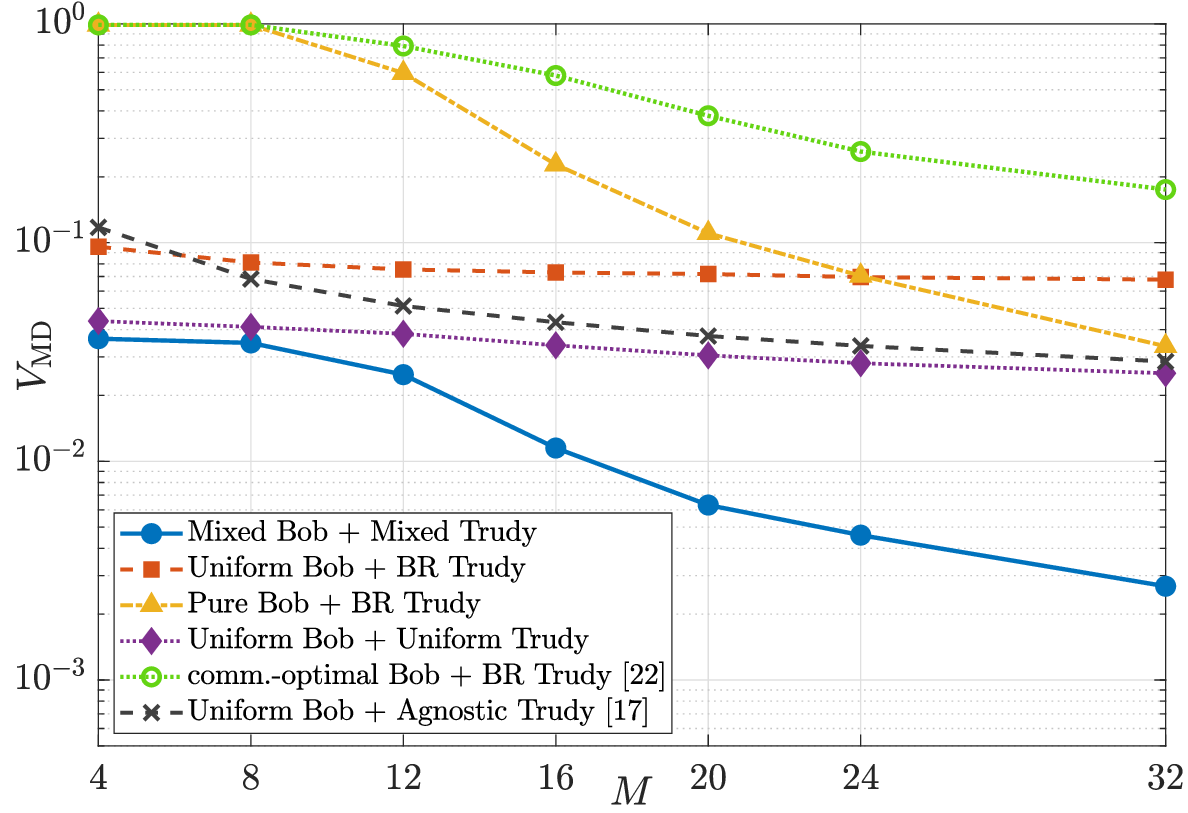}
\caption{
Comparison between the proposed minimax strategy and the considered baseline policies as a function of $M$, for $N_T=8$, $L_T=7$, $L_G=10$, $\mathrm{SNR}=5$~dB, and $P_{\mathrm{FA}}=10^{-2}$.
}
\label{fig:pmd_m_benchmark1}
\end{figure}

\begin{figure}[]
\centering
\includegraphics[width=\columnwidth]{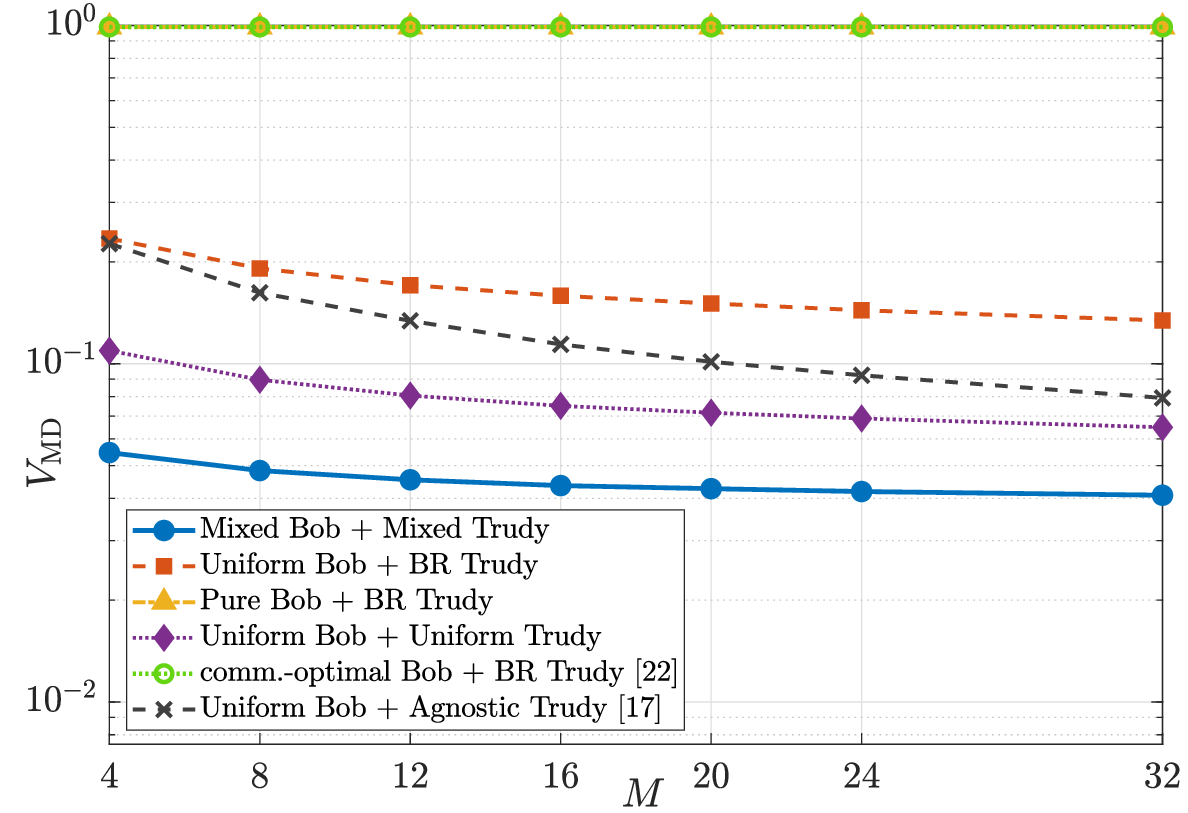}
\caption{
Comparison between the proposed minimax strategy and the considered baseline policies as a function of $M$, for $N_T=8$, $L_T=7$, $L_G=3$, $\mathrm{SNR}=5$~dB, and $P_{\mathrm{FA}}=10^{-2}$.
}
\label{fig:pmd_m_benchmark2}
\end{figure}

Figs.~\ref{fig:pmd_m_benchmark1} and~\ref{fig:pmd_m_benchmark2} compare the equilibrium value obtained with the proposed mixed-strategy zero-sum game against the five baselines introduced above. The parameters are $N_T=8$, $L_T=7$, $\mathrm{SNR}=5$~dB, and $P_{\mathrm{FA}}=10^{-2}$. The two figures consider $L_G=10$ and $L_G=3$, respectively.

Given Eq. \eqref{eq:Trudy_cond}, the configuration-agnostic baseline is not guaranteed to attain the perfect indistinguishability limit for this parameter setting. Section~\ref{sec:3gpp_perfect} instead considers a full-controllability regime in which this limit is recovered.

For high spatial richness ($L_G=10$), Trudy becomes increasingly spatially constrained as Bob's receive array $M$ grows. When $M$ is small, \eqref{eq:omniscient_generic_simplified} is satisfied and Trudy can match the effective observation dimension, allowing pure-Bob strategies to approach the indistinguishability limit. As $M$ increases, Bob exploits a higher-dimensional channel space while Trudy's effective rank remains capped at $7$, progressively reducing attack performance. Furthermore, optimizing RIS configurations exclusively for communication performance incurs severe security degradation compared to security-oriented designs, yielding a substantially higher miss-detection probability $V_{\mathrm{MD}}$.

The configuration-agnostic attack also becomes less effective with larger $M$ because the rank-deficient matrix $\bm T$ leaves residual channel mismatches that Bob's array can resolve. Nonetheless, it outperforms uniform randomization over Trudy's discrete codebook $\mathcal C_Q$ by explicitly minimizing pre-RIS mismatch. 
 Since $\bm Q_{\rm agn}$ is included in Trudy's codebook, her best response over $\mathcal C_Q$ performs at least as well as this individual precoder for any fixed strategy of Bob. While uniform RIS randomization at Bob improves resilience over deterministic policies, it remains suboptimal. The proposed minimax policy optimizes the full probability distribution, consistently achieving the lowest $V_{\mathrm{MD}}$.

Under low spatial richness ($L_G=3$), Trudy possesses sufficient spatial degrees of freedom ($\min\{N_T,L_T\}=7 > L_G$) to reproduce the effective channel for any individually selected RIS configuration. Consequently, pure-Bob and communication-optimal baselines remain pinned at the indistinguishability limit regardless of $M$. The configuration-agnostic attack still stays below this limit because a static precoder cannot simultaneously match all DFT configurations.

Finally, evaluating the reciprocal benchmark where Trudy commits to a precoder in $\mathcal C_Q$ and Bob best-responds yields numerically negligible $V_{\mathrm{MD}}$ as $M$ grows. This confirms that no single precoder in $\mathcal C_Q$ is universally robust, proving the absence of a pure-strategy saddle point and validating mixed randomization strategies.

\paragraph*{Effect Of The RIS Codebook Design}

\begin{figure}[]
\centering
\includegraphics[width=\columnwidth]{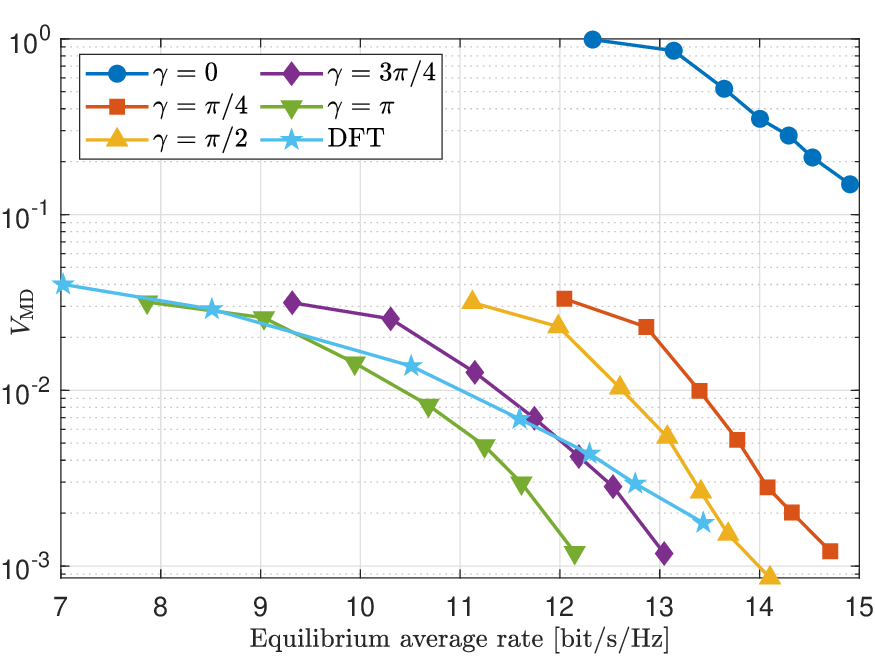}
\caption{
$V_{\rm MD}$ as a function of the equilibrium-weighted average legitimate rate for the DFT-based RIS codebook and codebooks uniformly perturbed around the communication-optimal configuration, for $\gamma\in{0,\pi/4,\pi/2,3\pi/4,\pi}$, $N_T=16$, $L_T=5$, $L_G=7$, $\mathrm{SNR}=5$~dB, and $P_{\rm FA}=10^{-2}$. The points along each curve correspond to $M\in{4,8,12,16,20,24,32}$.
}
\label{fig:pmd_rate_tradeoff}
\end{figure}

To investigate how the RIS codebook design affects both authentication and communication performance, we consider an alternative construction inspired by~\cite{tomasin2024analysis} that generates the codebook around the communication-optimal configuration $\overline{\bm\omega}$.~\footnote{Several works address RIS configuration optimization; see the survey~\cite{10361836} and the recent contributions~\cite{Guglielmi1,10417011,globecom2023}.} It therefore provides a tunable compromise between preserving the legitimate-channel rate and introducing the challenge diversity required by CR-PLA. The $k$-th codeword is generated as
\begin{equation}
\bm\omega_k(\gamma)
=
\overline{\bm\omega}
\odot
\exp\left(j\bm\delta_k\right),
\qquad
[\bm\delta_k]_n\sim\mathcal U[-\gamma,\gamma],
\end{equation}
where the phase perturbations are independently generated across RIS elements and codewords, $\gamma\in[0,\pi]$ controls the departure from $\overline{\bm\omega}$, and $\odot$ denotes the Hadamard product.

Fig.~\ref{fig:pmd_rate_tradeoff} reports the tradeoff between the equilibrium-weighted average legitimate rate and $V_{\rm MD}$ for $\gamma\in\{0,\pi/4,\pi/2,3\pi/4,\pi\}$.  For $K=1$ and unit transmit power, the legitimate rate follows from the Shannon formula as $R(\bm\omega)=\log_2(1+\mathrm{SNR} \|\bm G\operatorname{diag}(\bm\omega)\bm F\|^2)$. The equilibrium-weighted rate is then $\sum_{i=1}^{N_\Omega}o_i^\star R(\bm\omega_i)$, where $o_i^\star$ is Bob's equilibrium probability. 
For $\gamma=0$, all the RIS codewords collapse to $\overline{\bm\omega}$. Bob therefore loses challenge diversity, and Trudy's configuration-dependent attack remains highly effective. This design provides the highest communication rate but yields $V_{\rm MD}$ close to the indistinguishability limit.


Increasing perturbation $\gamma$ significantly reduces $V_{\rm MD}$ by rendering targeted precoders ineffective, though at the expense of lower legitimate rates. Intermediate values ($\gamma=\pi/4, \pi/2$) strike an effective compromise, capturing most security gains with minimal rate loss. Larger perturbations saturate non-monotonically, as game value depends on full codebook geometry and equilibrium probabilities rather than maximum phase displacement alone.

At $\gamma=\pi$, every phase is uniformly distributed over the unit circle, so the codebook no longer retains a statistical preference for $\overline{\bm\omega}$. Nevertheless, it does not coincide with the DFT codebook. The DFT vectors are deterministically orthogonal, whereas independently generated random vectors are orthogonal only in expectation. Defining the mutual coherence as
\begin{equation}
\mu = 
\max_{i\neq j}
\frac{\left|\langle\bm\omega_i,\bm\omega_j\rangle\right|}{N},
\end{equation}
the DFT codebook satisfies $\mu=0$ up to numerical precision, whereas the considered random $\gamma=\pi$ codebooks exhibit an average coherence of approximately $0.44$ for $N=32$.

For the smallest arrays, the relative ordering of the two designs depends on the operating point. For intermediate values of $M$, the DFT-based codebook provides lower $V_{\rm MD}$ together with a higher rate, while for the largest arrays both designs yield extremely small and eventually numerically indistinguishable \acp{mdp}. The DFT design nevertheless retains an equilibrium-rate advantage of approximately $1.5$~bit/s/Hz.

Therefore, increasing $\gamma$ improves security by sacrificing part of the communication gain, but full-range randomization is not equivalent to a DFT codebook. The deterministic orthogonality of the latter may provide an additional advantage when Bob can exploit the resulting codebook geometry.

\subsection{Performance on the 3GPP Channel}
\label{sec:3gpp_perfect}

\begin{figure}[]
\centering
\includegraphics[width=\columnwidth]{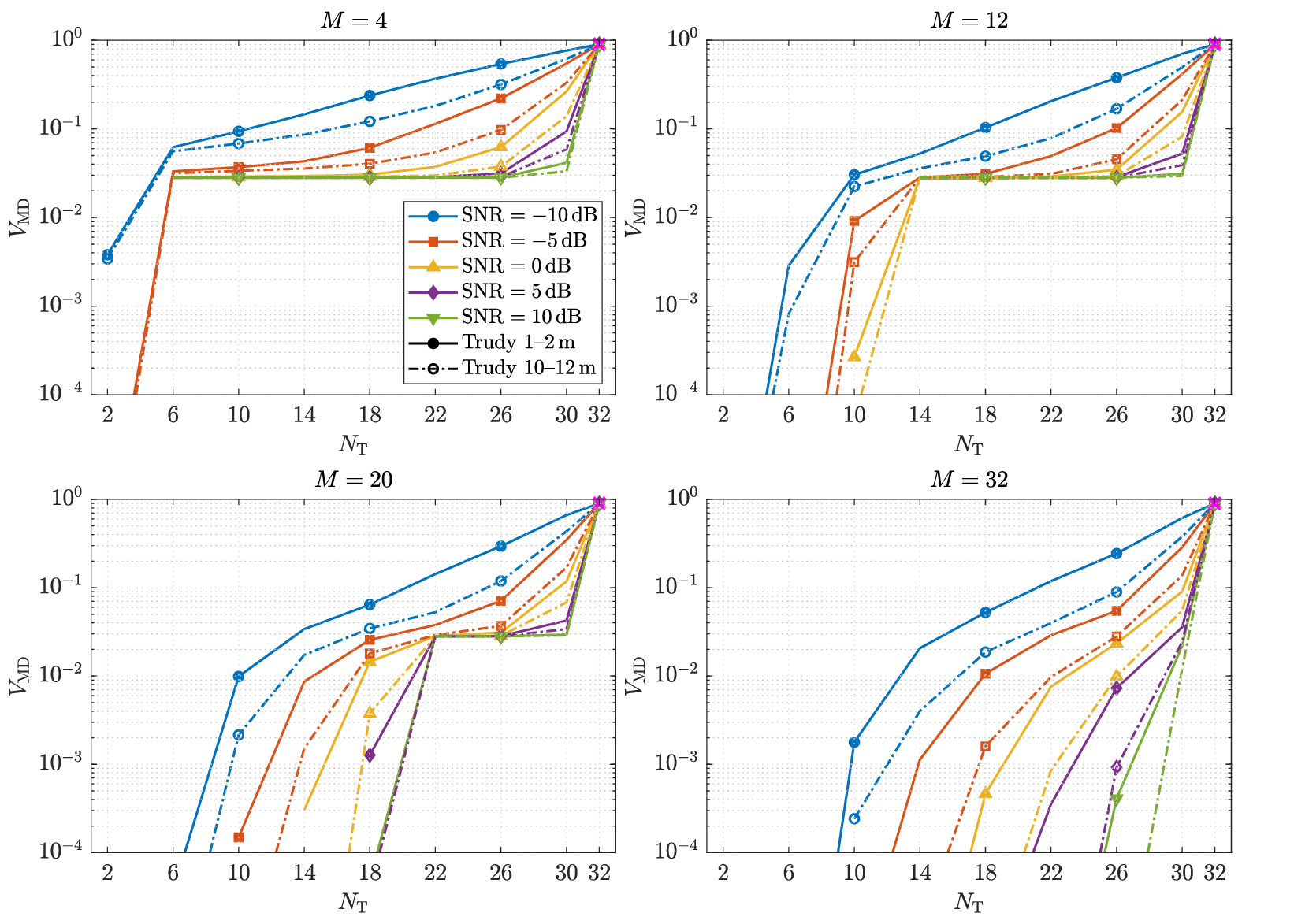}
\caption{
$V_{\mathrm{MD}}$ as a function of $N_T$, for
$P_{\mathrm{FA}}=10^{-1}$.
The four panels correspond to $M\in\{4,12,20,32\}$.
Colors and markers identify the considered SNR values,
while solid lines with filled markers and
dash-dotted lines with empty markers refer to the
Trudy--Alice distance ranges $1$--$2$~m and $10$--$12$~m,
respectively. 
The magenta cross at $N_T=32$ in each panel marks the
indistinguishability limit $V_{\mathrm{MD}}=1-P_{\mathrm{FA}}$
attained by the configuration-agnostic perfect through-RIS
attack when $\bm T$ is nonsingular.}

\label{fig:pmd_nt_pfa1}
\end{figure}

We evaluate the proposed framework in a 3GPP TR~38.901 UMi-NLOS propagation environment at 28~GHz with a single-antenna Alice and an N=32-element RIS. Narrowband channel matrices generated via QuaDRiGa are power-normalized, isolating spatial and multipath structural differences across two Trudy--Alice distance intervals (1--2~m and 10--12~m) from pure path loss. The simulation evaluates Bob antenna setups M{4,12,20,32}, Trudy antenna numbers $N_{\rm T}\in\{2,6,10,14,18,22,26,30,32\}$, SNR values in {-10,-5,0,5,10}~dB, and target false-alarm probability PFA=10-1. Trudy operates without transmit-power constraints, using a codebook containing both configuration-dependent precoders and the configuration-agnostic precoder $\bm Q_{\rm agn}$.
As shown in Fig.~\ref{fig:pmd_nt_pfa1}, authentication performance VMD is governed by the relative spatial degrees of freedom between Bob (M) and Trudy (NT). For NT<N, increasing M enhances Bob's ability to resolve residual mismatches between legitimate and forged channels. At low SNR, channel-estimation noise masks residual mismatches, resulting in higher VMD. Raising SNR improves Bob's discrimination capability when Trudy lacks spatial dimensions, but yields minimal gain when she can replicate the legitimate channel almost exactly.
Terminal separation also affects attack efficacy due to 3GPP spatial consistency. A nearby Trudy shares strong cluster-level channel features with Alice, whereas a more distant Trudy (10--12~m) requires a larger NT to achieve the same attack effectiveness as a nearby adversary (1--2~m).
Finally, a perfect indistinguishability regime emerges at NT=N=32 for all M. When $\bm T$ is nonsingular, the configuration-agnostic precoder $\bm Q_{\rm agn} = \bm T^{-1}\bm F$ satisfies $\bm G\bm\Omega_i\bm T\bm Q_{\rm agn} = \bm G\bm\Omega_i\bm F$ for every RIS configuration i. Under unconstrained transmit power, forged and legitimate responses coincide, yielding VMD=1-PFA independently of M and SNR. This recovers the perfect through-RIS attack limit of~\cite{tomasin2024analysis}, marked by the magenta crosses in Fig.~\ref{fig:pmd_nt_pfa1}.

\section{Conclusions}
\label{sec:conclusions}

In Part I of this two-part paper, we investigated RIS-assisted CR-PLA under perfect channel knowledge. We derived the conditions under which a multi-antenna attacker can exactly reproduce the legitimate cascaded channel and formulated the interaction between the verifier and the attacker as a finite zero-sum game. The resulting equilibrium determines Bob's optimal randomized RIS policy, Trudy's worst-case precoding strategy, and the corresponding equilibrium \ac{mdp}.

Numerical results based on geometric multipath and 3GPP UMi-NLOS channel models showed that authentication performance is mainly governed by the spatial degrees of freedom available to Bob and Trudy. Optimized RIS randomization provides substantial security gains whenever no single precoder can reproduce Alice's channel across the entire RIS codebook. Conversely, when a configuration-agnostic attack is feasible, RIS randomization loses its advantage and \ac{mdp} reaches $1-P_{\mathrm{FA}}$. The results also highlighted the rate--security tradeoff induced by the RIS codebook design. 


\balance
\bibliographystyle{IEEEtran}
\bibliography{biblio2}

\end{document}